\documentclass[%
aps,
pre,
superscriptaddress,
tightenlines,
showpacs,showkeys,
a4paper,
12pt,
notitlepage
]{revtex4-1}
\usepackage[english]{babel}
\usepackage{amssymb,amsmath,stmaryrd,array}

\usepackage{graphicx}
\usepackage{subfig}

\makeatletter

\newcommand{\sca}[2]{\ensuremath{\bigl({#1}\cdot{#2}\bigr)}}
\newcommand{\avr}[1]{\ensuremath{\langle{#1}\rangle}}

\newcommand{\cnj}[1]{{#1}^{\ast}}
\newcommand{\hcnj}[1]{{#1}^{\dagger}}
\newcommand{\tcnj}[1]{{#1}^{T}}

\newcommand{\pdrs}[1]{\partial_{#1}}
\newcommand{\pdr}[2]{\frac{\partial #1}{\partial #2}}

\newcommand{\diag}{\mathop{\rm diag}\nolimits}
\newcommand{\sign}{\mathop{\rm sign}\nolimits}

\newcommand{\Tr}{\mathop{\rm Tr}\nolimits}

\newcommand{\mum}{$\mu$m}
 \newcommand{\bs}[1]{\boldsymbol{#1}}
 \newcommand{\vc}[1]{\mathbf{#1}}
 \newcommand{\mvc}[1]{\mathbf{#1}}
 \newcommand{\uvc}[1]{\hat{\mathbf{#1}}}
 
 \newcommand{\ind}[1]{\mathrm{#1}}

\newcommand{\dd}{\mathrm{d}}

\newcommand{\eff}{\mathrm{eff}}

\newcommand{\inc}{\mathrm{inc}}
\newcommand{\refl}{\mathrm{refl}}
\newcommand{\transm}{\mathrm{trm}}
\newcommand{\vac}{\mathrm{vac}}
\newcommand{\med}{\mathrm{m}}

\makeatother

\begin{document}
\DeclareGraphicsExtensions{.eps,.png,.pdf}
\title{
Remarkable optics of
short-pitch deformed helix ferroelectric liquid crystals:
symmetries, exceptional points 
and polarization-resolved angular patterns  
}

\author{Alexei~D.~Kiselev}
\email[Email address: ]{kiselev@iop.kiev.ua}
\affiliation{%
 Hong Kong University of Science and Technology,
 Clear Water Bay, Kowloon, Hong Kong
 }
\affiliation{%
 Institute of Physics of National Academy of Sciences of Ukraine,
 prospekt Nauki 46,
 03680 Kiev, Ukraine}

 \author{Vladimir~G.~Chigrinov}
 \email[Email address: ]{eechigr@ust.hk}
\affiliation{%
 Hong Kong University of Science and Technology,
 Clear Water Bay, Kowloon, Hong Kong
 }

\date{\today}

\begin{abstract}
In order to explore 
electric-field-induced transformations
of polarization singularities in
the polarization-resolved angular (conoscopic)
patterns emerging after
deformed helix ferroelectric liquid crystal (DHFLC) cells
with subwavelength helix pitch,
we combine the transfer
matrix formalism with the results for the effective
dielectric tensor of biaxial FLCs evaluated using
an improved technique of averaging
over distorted helical structures.
Within the framework of the transfer matrix method,
we deduce a number of symmetry relations
and show that the symmetry axis of $L$
lines (curves of linear polarization)
is directed along the major in-plane optical axis
which rotates under the action of the electric field.
When the angle between this axis and
the polarization plane of incident linearly polarized light
is above its critical value,
the $C$ points
(points of circular polarization)
appear in the form of
symmetrically arranged chains of densely packed 
star-monstar pairs.
We also emphasize the role of phase singularities of a different kind
and discuss the enhanced electro-optic response of DHFLCs
near the exceptional point where the condition of zero-field
isotropy is fulfilled.
\end{abstract}

 \maketitle

%%%%%%%%%%%%%%
\section{Introduction}
\label{sec:intro}
%%%%%%%%%%%%%%

\begin{figure*}[!tbh]
\centering
\subfloat[Smectic cone]{
  %\resizebox{65mm}{!}{\includegraphics*{smectic-cone.eps}}
  \resizebox{65mm}{!}{\includegraphics*{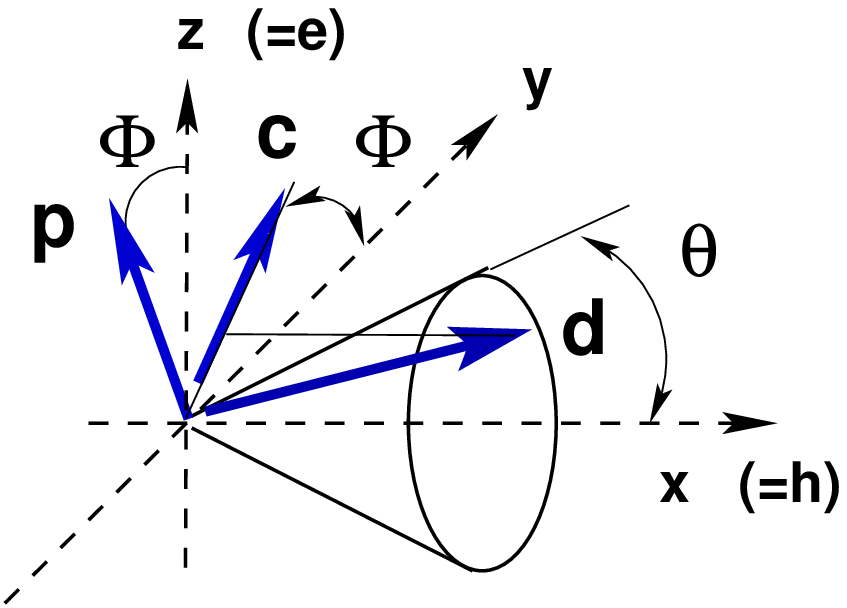}}
\label{subfig:director}
}
\subfloat[Planar aligned FLC cell]{
  %\resizebox{90mm}{!}{\includegraphics*{dhf-cell.eps}}
  \resizebox{65mm}{!}{\includegraphics*{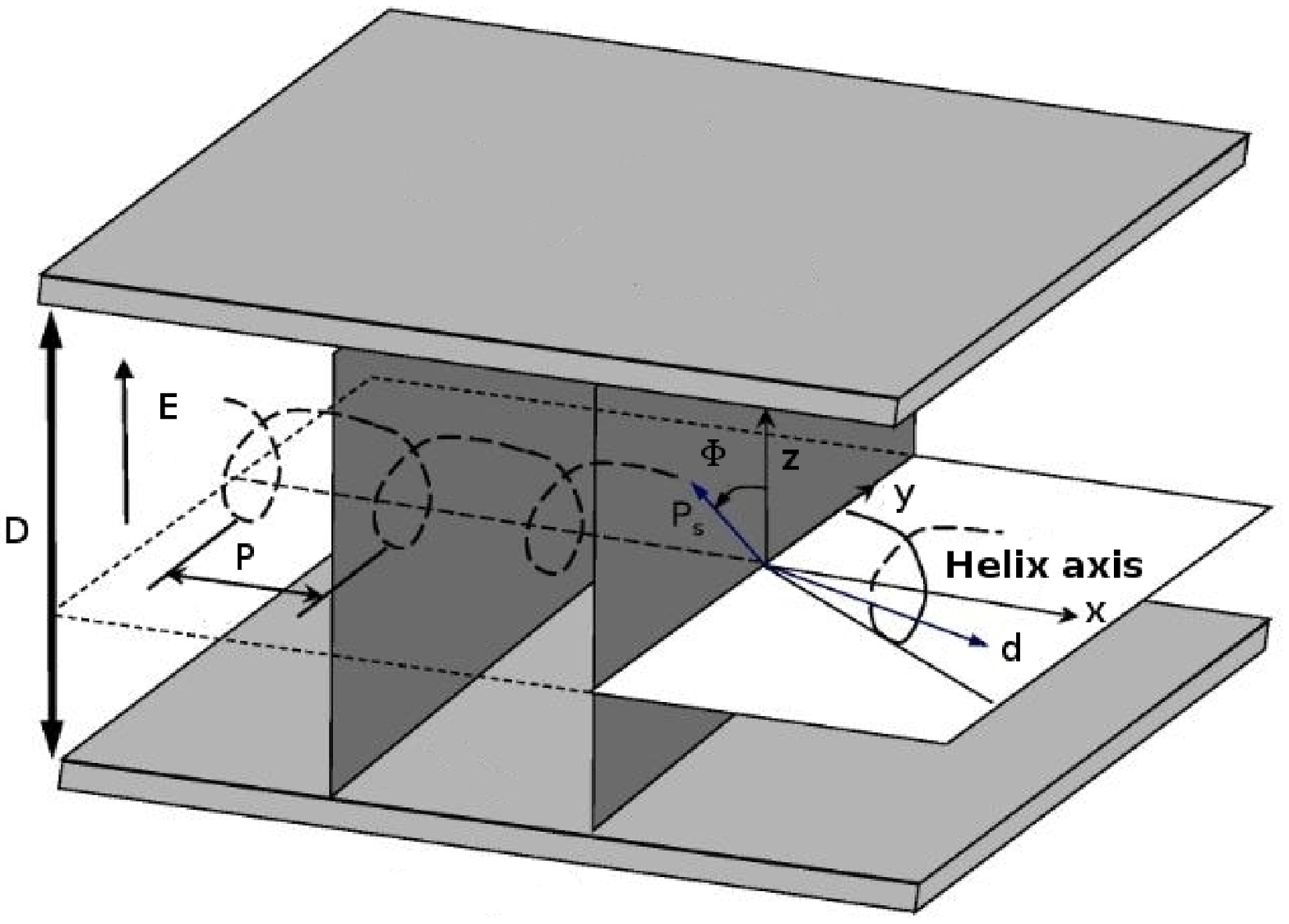}}
\label{subfig:cell}
}
\caption{%
(Color online)
  Geometry of (a)~smectic cone and (b)~planar aligned FLC cell with
uniform lying helix. 
}
\label{fig:geom}
\end{figure*}

Over the last more than three decades ferroelectric liquid crystals
(FLCs)
have attracted considerable attention
as promising chiral liquid crystal materials for applications in 
fast switching display devices
(a detailed description of FLCs can be found, e.g.,
in monographs~\cite{Lagerwall:bk:1999,Oswald:bk:2006}).
Equilibrium orientational structures in FLCs
are represented by helical twisting patterns
where FLC molecules align on average along
a local unit director
\begin{align}
&
\uvc{d}=
\cos\theta\,\uvc{h}+
\sin\theta\,\uvc{c},
\label{eq:director}
  \end{align}
where $\theta$ is the smectic tilt angle; 
$\uvc{h}$ is the twisting axis normal to the smectic layers and
$\uvc{c}\perp\uvc{h}$ is the $c$-director.
The FLC director~\eqref{eq:director}
lies on the smectic cone 
depicted in Fig.~\ref{subfig:director}
with
the \textit{smectic tilt angle} $\theta$
and rotates
in a helical fashion about a uniform twisting axis
$\uvc{h}$ forming the FLC helix with 
the \textit{helix pitch}, $P$.
This rotation is described by
the azimuthal angle
around the cone $\Phi$
that specifies
orientation of the $c$-director in the plane perpendicular to
$\uvc{h}$ and depends on
the dimensionless coordinate along the twisting axis
\begin{align}
&
\phi=2 \pi \sca{\uvc{h}}{\vc{r}}/P= q x,
\label{eq:phi}    
 \end{align}
where $q=2\pi/P$ is the helix twist wave number. 

The important case of a uniform lying FLC helix
in the slab geometry 
with the smectic layers normal to the substrates
and
\begin{align}
&
\uvc{h}=\uvc{x},
\quad
\uvc{c}=\cos\Phi\,\uvc{y}+
\sin\Phi\,\uvc{z},
\quad
\vc{E}=E\,\uvc{z},
\label{eq:director-planar}    
  \end{align}
where $\vc{E}$ is the electric field applied across the cell,
is illustrated in Fig.~\ref{fig:geom}. 
This  is the geometry of surface stabilized FLCs
(SSFLCs)
pioneered by Clark and Lagerwall in~\cite{Clark:apl:1980}.
They studied electro-optic response of 
FLC cells confined between two parallel plates
subject to homogeneous boundary conditions
and made thin enough to suppress the bulk FLC 
helix. 

It was found that
such cells exhibit high-speed, bistable electro-optical switching
between orientational states stabilized by surface interactions.
The response of FLCs
to an applied electric field $\vc{E}$
is characterized by fast switching times due to
linear coupling between the field and 
the \textit{spontaneous ferroelectric polarization}
\begin{align}
&
\vc{P}_s=P_s\uvc{p},
\quad
\uvc{p}=
\uvc{h}\times\uvc{c}=\cos\Phi\,\uvc{z}-
\sin\Phi\,\uvc{y},
\label{eq:pol-vector}
  \end{align}
where $\uvc{p}$ is
the \textit{polarization unit vector}.
There is also a threshold voltage necessary for switching to occur
and the process of bistable switching is typically accompanied
by a hysteresis.

Figure~\ref{subfig:cell} also describes
the geometry of 
deformed helix FLCs (DHFLCs)
as it was introduced in~\cite{Beresnev:lc:1989}.
This case will be of our primary concern.

In DHFLC cells, the FLC helix
is characterized by
a short submicron \textit{helix pitch}, $P<1$~\mum, 
and a relatively large \textit{tilt angle}, 
$\theta>30$~deg. 
By contrast to SSFLC cells,
where the surface induced unwinding
of the bulk helix requires
the helix pitch of a FLC mixture to be greater than
the cell thickness,
a DHFLC helix pitch is 5-10 times smaller than
the thickness. This allows the helix to be retained within
the cell boundaries. 

Electro-optical response of DHFLC cells
exhibits a  number of peculiarities that make them
useful for LC devices such as
high speed spatial light 
modulators~\cite{Abdul:mclc:1991,Cohen:aplopt:1997,Pozhidaev:ferro:2000,Kiselev:pre:2013,Kiselev:ol:2014}, 
colour-sequential liquid crystal display cells~\cite{Hedge:lc:2008}
and optic fiber sensors~\cite{Kiselev:photsens:2012}.
The effects caused by electric-field-induced distortions of the helical structure 
underline the mode of operation of such cells. 
In a typical experimental setup, these effects
are probed by performing measurements of 
the transmittance of normally incident linearly polarized light
through a cell placed between crossed polarizers.

A more general case of oblique incidence
has not received a fair amount of attention.
Theoretically, a powerful tool to deal with 
this case is the transfer matrix method
which has been widely used
in studies of both quantum mechanical and optical wave 
fields~\cite{Markos:bk:2008,Yariv:bk:2007}.
In this work  
we apply the method for systematic treatment of
the technologically important case of 
DHFLCs with subwavelength pitch also known as 
the \textit{short-pitch DHFLCs}.

Recently, the transfer matrix approach to polarization
gratings was employed to define
the effective dielectric tensor of short-pitch 
DHFLCs~\cite{Kiselev:pre:2011}
that gives the principal values and orientation of the optical axes
as a function of the applied electric field.
Biaxial anisotropy and rotation of the in-plane optical axes
produced by the electric field
can be interpreted as 
the \textit{orientational Kerr effect}~\cite{Kiselev:pre:2013,Kiselev:ol:2014}.

It can be expected that the electric field dependence of the effective dielectric tensor
will also manifest itself as electric-field-induced transformations
of the polarization-resolved angular (conoscopic)  patterns
in the observation plane after the DHFLC cells illuminated by 
convergent light beam.
These patterns are represented by 
the two-dimensional (2D) fields of polarization 
ellipses describing the polarization structure behind the
conoscopic images~\cite{Kiselev:jpcm:2007,Kiselev:pra:2008}.

As it was originally recognized by
Nye~\cite{Nye:prsl:1983a,Nye:prsl:1987,Nye:bk:1999},
the key elements characterizing 
geometry of such Stokes parameter fields
are the \textit{polarization singularities}
that play the fundamentally important role of structurally
stable topological defects
(a recent review can be found in Ref.~\cite{Dennis:progr_opt:2009}).
In particular,
the polarization singularities
such as the \textit{C points}
(the points where the light wave is circular polarized)
and the \textit{L lines}
(the curves along which the polarization is linear)
frequently emerge as the characteristic feature
of certain polarization state distributions.
For nematic and 
cholescteric (chiral nematic) liquid crystals,
the singularity structure of the polarization-resolved angular
patterns is generally found to be sensitive to
both the director configuration and the polarization characteristics of
incident light~\cite{Kiselev:jpcm:2007,Kiselev:pra:2008,Egorov:apb:2010}.

In this study, we consider 
the polarization-resolved angular
patterns of DHFLC cells 
as the Stokes parameter fields 
giving detailed information
on the incidence angles dependence of the polarization state of
light transmitted through the cells.
In particular, we explore how
the polarization singularities transform under the action of the
electric field. Our analysis will utilize the transfer
matrix approach in combination with the results for the effective
dielectric tensor  of biaxial FLCs evaluated using
an improved technique of averaging
over distorted helical structures.
We also emphasize the role of phase singularities of a different kind
and discuss the electro-optic behavior of DHFLCs
near the exceptional point where the condition of zero-field
isotropy is fulfilled.
 
The layout of the paper is as follows.
In Sec.~\ref{sec:theory} we introduce our notations and
describe the transfer matrix formalism rendered into the $4\times 4$
matrix form suitable for our purposes.
This formalism is employed to deduce a number of the unitarity
and symmetry relations with emphasis on the planar anisotropic
structures that represent DHFLC cells and posses
two optical axes lying in the plane of substrates.
In Sec.~\ref{sec:optics-dhf}
we evaluate the effective dielectric tensor of DHFLC cells, 
discuss the orientational Kerr
effect and show that electro-optic response of DHFLC cells is enhanced
near the exceptional point determined by the condition of zero-field isotropy.  
Geometry of the polarization-resolved angular patterns
emerging after DHFLC cells
is considered in Sec.~\ref{sec:conoscopy}.
After providing necessary details  on 
our computational approach and the polarization singularities, 
we present the numerical results describing 
how the singularity structure of polarization ellipse fields
transforms under the action of the electric field.
Finally, in Sec.~\ref{sec:discussion} we draw the results together and
make some concluding remarks.
Details on some technical results are relegated to Appendixes~\ref{sec:op-evol}--\ref{sec:FLC-spiral}.

\begin{figure*}[!tbh]
\centering
  %\resizebox{90mm}{!}{\includegraphics*{transfer_4rays.eps}}
\resizebox{90mm}{!}{\includegraphics*{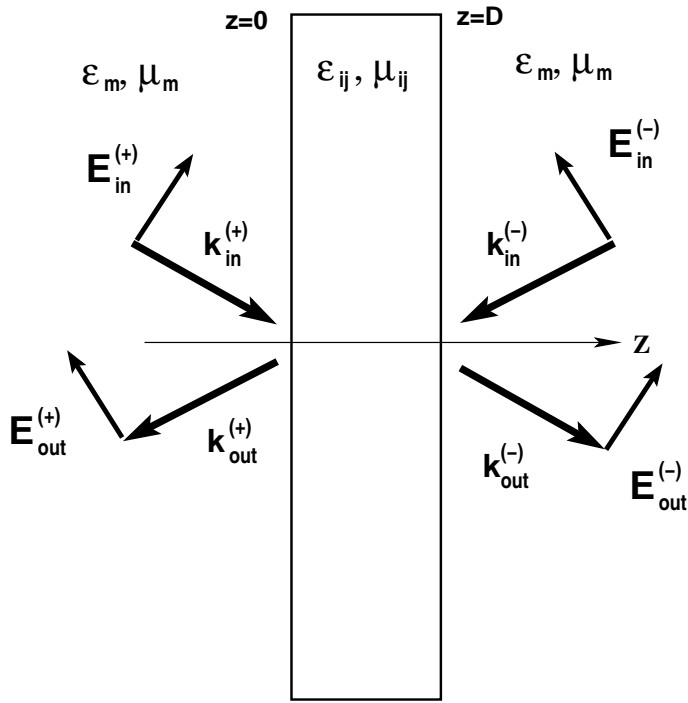}}
\caption{%
Four-wave geometry with two incoming (incident) waves,
$\vc{E}_{\ind{in}}^{(+)}$ and $\vc{E}_{\ind{in}}^{(-)}$,
impinging onto the entrance ($z=0$) and exit ($z=D$) 
faces, respectively.
}
\label{fig:geom_4rays}
\end{figure*}

%%%%%%%%%%%%%%%%%%%%%%%%%%
\section{Transfer matrix method and symmetries}
\label{sec:theory}
%%%%%%%%%%%%%%%%%%%%%%%%%%

In order to describe 
both
the electro-optical properties and the polarization-resolved
angular patterns 
of deformed helix ferroelectric liquid crystal layers
with subwavelength pitch
we adapt a general theoretical approach
which can be regarded as a modified version of 
the well-known transfer matrix method~\cite{Markos:bk:2008,Yariv:bk:2007}
and was previously applied to study the polarization-resolved conoscopic
patterns of nematic liquid crystal 
cells~\cite{Kiselev:jpcm:2007,Kiselev:pra:2008,Kiselev:jetp:2010}.
This approach has also been extended to the case of polarization
gratings and used to deduce the general expression for the effective
dielectric tensor of DHFLC cells~\cite{Kiselev:pre:2011}. 

In this section, 
we present the transfer matrix approach
as the starting point of our theoretical
considerations,
with emphasis on
its general structure and the symmetry relations.
The analytical results for uniformly anisotropic planar
structures
representing homogenized DHFLC cells
are given in Appendix~\ref{sec:planar}.  

We deal with a harmonic electromagnetic field 
characterized by the free-space wave number
$k_{\vac}=\omega/c$,
where $\omega$ is the frequency
(time-dependent factor
is $\exp\{-\omega t\}$),
and consider the slab geometry shown in Fig.~\ref{fig:geom_4rays}. 
In this geometry, an optically anisotropic layer
of thickness $D$
is sandwiched between 
the bounding surfaces (substrates): $z=0$ and $z=D$
(the $z$ axis is normal to the substrates)
and is
characterized by the dielectric tensor $\epsilon_{ij}$
and the magnetic permittivity $\mu$

Further, 
we restrict ourselves to the case of stratified media
and assume that
the electromagnetic fields can be taken in the following
factorized form
\begin{align}
  \label{eq:EH-form}
\{\vc{E}(\vc{r}), \vc{H}(\vc{r})\}=
\{\vc{E}(z), \vc{H}(z)\}\exp\sca{\vc{k}_p}{\vc{r}},
\end{align}
where the vector
\begin{align}
  \label{eq:k_p}
  \vc{k}_p/k_{\vac}=\vc{q}_p=(q_x^{(p)},q_y^{(p)},0)=
q_p(\cos\phi_p,\sin\phi_p,0)
\end{align}
represents the lateral component of the wave vector.
Then we write down
the representation for the electric and magnetic fields, $\vc{E}$ and
$\vc{H}$,
\begin{align}
  \label{eq:decomp-E}
  \vc{E}=E_z \uvc{z} +\vc{E}_{P},\quad
\vc{H}= H_z \uvc{z} +\uvc{z}\times \vc{H}_{P},
\end{align}
where the  components directed along the normal to the bounding surface
(the $z$ axis) are separated from the tangential (lateral) ones. 
In this representation,
the vectors
$\vc{E}_{P}=E_x \uvc{x}+E_y \uvc{y}\equiv
\begin{pmatrix}
  E_x\\E_y
\end{pmatrix}
$
and
$\vc{H}_{P}=\vc{H}\times\uvc{z}\equiv
\begin{pmatrix}
  H_y\\-H_x
\end{pmatrix}
$
are parallel to the substrates
and give the lateral components of the electromagnetic field.

Substituting the relations~\eqref{eq:decomp-E}
into the Maxwell equations and
eliminating the $z$ components of the electric and 
magnetic fields gives
equations for 
the tangential components of the electromagnetic field
that can be written  
in the following $4\times 4$ matrix 
form~\cite{Kiselev:pra:2008,Kiselev:pre:2011}:
\begin{align}
  \label{eq:matrix-system}
  -i\pdrs{\tau}\vc{F}=\mvc{M}\cdot\vc{F}\equiv
    \begin{pmatrix}\mvc{M}_{11}&\mvc{M}_{12}\\\mvc{M}_{21}&\mvc{M}_{22} \end{pmatrix}
    \begin{pmatrix}\vc{E}_{P}\\\vc{H}_{P} \end{pmatrix},
\quad
\tau\equiv k_{\vac} z,
\end{align}
where  
$\mvc{M}$ is the \textit{differential propagation matrix} 
 and
its $2\times 2$ block matrices
$\mvc{M}_{ij}$ are given by
\begin{subequations}
\label{eq:Mij}
   \begin{align}
&
\label{eq:Mii}
\mvc{M}^{(11)}_{\alpha\beta}=
-\epsilon_{zz}^{-1}
q_{\alpha}^{(p)}\epsilon_{z \beta},
\quad
\mvc{M}^{(22)}_{\alpha\beta}=-\epsilon_{zz}^{-1}\epsilon_{\alpha z}
q_{\beta}^{(p)},
\\
&
\label{eq:M12}
\mvc{M}^{(12)}_{\alpha\beta}=\mu 
\delta_{\alpha \beta}- 
q_{\alpha}^{(p)}\, \epsilon_{zz}^{-1}\,q_{\beta}^{(p)},
\\
&
\label{eq:M21}
\mvc{M}^{(21)}_{\alpha\beta}=\epsilon_{\alpha\beta}-
\epsilon_{\alpha z}\epsilon_{zz}^{-1}\epsilon_{z\beta}
-
\mu^{-1}p_{\alpha}^{(p)}p_{\beta}^{(p)},\quad
\vc{p}_p=\uvc{z}\times\vc{q}_p.     
\end{align}
\end{subequations}
% and $\eta_{zz}$,
% $\beta_{ij}$ and $\epsilon_{\alpha\beta}^{(P)}$
% are the Fourier coefficients for
% $\epsilon_{zz}^{-1}$,
% $\epsilon_{zz}^{-1}\epsilon_{ij}$
% and $\epsilon_{\alpha\beta}^{(P)}$, respectively.

General solution of the system~\eqref{eq:matrix-system}
\begin{align}
  \label{eq:evol-oprt}
  \vc{F}(\tau)=
  \mvc{U}(\tau,\tau_0)\cdot\vc{F}(\tau_0)
\end{align}
can be conveniently expressed in terms of
the \textit{evolution operator}
which is also known as the 
\textit{propagator} and is
defined as the matrix solution of 
the initial value problem
\begin{subequations}
  \label{eq:evol_problem}
\begin{align}
  \label{eq:evol_eq}
     -i\pdrs{\tau}\mvc{U}(\tau,\tau_0)
&
=
\mvc{M}(\tau)\cdot\mvc{U}(\tau,\tau_0),
\\
  \label{eq:evol_ic}
\mvc{U}(\tau_0,\tau_0)
&
=\mvc{I}_4,
\end{align}
\end{subequations}
where $\mvc{I}_n$ is the $n\times n$ identity matrix.
Basic properties of the evolution operator
are reviewed in Appendix~\ref{sec:op-evol}. 

%%%%%%%%%%%%%%%%%%%%%%
\subsection{Input-output  relations}
\label{subsec:input-out}
%%%%%%%%%%%%%%%%%%%%%

In the ambient medium with $\epsilon_{ij}=\epsilon_{\med}\delta_{ij}$
and $\mu=\mu_{\med}$, the general solution~\eqref{eq:evol-oprt}
can be expressed in terms of
plane waves propagating along  
the wave vectors with the tangential component~\eqref{eq:k_p}.
For such waves, the result  is given by
\begin{align}
&
  \label{eq:F_med}
  \vc{F}_{\med}(\tau)=\mvc{V}_{\med}(\vc{q}_p)
  \begin{pmatrix}
    \exp\{i \mvc{Q}_{\med}\, \tau\} & \mvc{0}\\
\mvc{0} &    \exp\{-i \mvc{Q}_{\med}\, \tau\}   
  \end{pmatrix}
  \begin{pmatrix}
    \vc{E}_{+}\\
\vc{E}_{-}
  \end{pmatrix},
\\
&
\label{eq:Q_med}
\mvc{Q}_{\med}=q_{\med}\,\mvc{I}_2,
\quad
q_{\med}=\sqrt{n_{\med}^2-q_p^2},
\end{align}
where 
$\mvc{V}_{\med}(\vc{q}_n)$
is the eigenvector matrix for the ambient medium
given by
\begin{align}
&
\label{eq:Vm-phi-q}
\mvc{V}_{\med}(\vc{q}_p)=
\mvc{T}_{\ind{rot}}(\phi_p)\mvc{V}_{\med}=
\begin{pmatrix}
 \mvc{Rt}(\phi_p)&\mvc{0}\\
\mvc{0}& \mvc{Rt}(\phi_p) 
\end{pmatrix}
\begin{pmatrix}
\mvc{E}_{\med} & -\bs{\sigma}_3 \mvc{E}_{\med}\\
\mvc{H}_{\med} & \bs{\sigma}_3 \mvc{H}_{\med}\\
\end{pmatrix},
\\
&
  \label{eq:EH-med}
  \mvc{E}_{\med}=
  \begin{pmatrix}
    q_{\med}/n_{\med}& 0\\
0 & 1
  \end{pmatrix},
\quad
 \mu_{\med}\,\mvc{H}_{\med}=
 \begin{pmatrix}
   n_{\med}& 0\\
0 & q_{\med}
 \end{pmatrix},
\\
&
\label{eq:Rot_matrix}
\mvc{Rt}(\phi)=\begin{pmatrix}
  \cos\phi &-\sin\phi\\
\sin\phi & \cos\phi
\end{pmatrix},
\end{align}
$\{\bs{\sigma}_1,\bs{\sigma}_2,\bs{\sigma}_3\}$ are the Pauli matrices
\begin{align}
  \label{eq:pauli}
      \bs{\sigma}_1=
      \begin{pmatrix}
        0&1\\1&0
      \end{pmatrix},
\:
      \bs{\sigma}_2=
      \begin{pmatrix}
        0&-i\\i&0
      \end{pmatrix},
\:
      \bs{\sigma}_3=
      \begin{pmatrix}
        1&0\\0&-1
      \end{pmatrix}.
\end{align}

From Eq.~\eqref{eq:F_med},
the vector  amplitudes $\vc{E}_{+}$ and
$\vc{E}_{-}$ correspond to the forward and backward eigenwaves
with
$\vc{k}_{+}=k_{\vac}(q_{\med}\,\uvc{z}+\vc{q}_p)$
and 
$\vc{k}_{-}=k_{\vac}(-q_{\med}\,\uvc{z}+\vc{q}_p)$, respectively.
Figure~\ref{fig:geom_4rays} shows that,
in the half space $z\le 0$ 
before the entrance face of the layer $z=0$,
these eigenwaves describe 
the \textit{incoming and outgoing waves}
 \begin{align}
   &
   \label{eq:in-out-plus}
   \vc{E}_{+}\vert_{z\le 0}=
\vc{E}_{\ind{in}}^{(+)},
\quad
   \vc{E}_{-}\vert_{z\le 0}=
   \vc{E}_{\ind{out}}^{(+)},
 \end{align} 
whereas,
in the half space $z\ge D$ after the exit face of the layer,
these waves are given by
\begin{align}
   \label{eq:in-out-minus}
   \vc{E}_{+}\vert_{z\ge D}=
\vc{E}_{\ind{out}}^{(-)},
\quad
   \vc{E}_{-}\vert_{z\ge D}=
   \vc{E}_{\ind{in}}^{(-)}.
\end{align}
In this geometry, there are two plane waves, 
$\vc{E}_{\ind{in}}^{(+)}$ and $\vc{E}_{\ind{in}}^{(-)}$,
incident on the bounding surfaces of the anisotropic layer,
$z=0$ and $z=D$, respectively.
Then the standard linear input-output relations
\begin{align}
&
  \label{eq:transm-rel}
\vc{E}_{\transm}
=\mvc{T}
\vc{E}_{\inc},
\quad
\vc{E}_{\refl}
=
\mvc{R}
\vc{E}_{\inc}
\end{align}
linking the vector amplitudes of transmitted and reflected waves,
$\vc{E}_{\transm}$ and $\vc{E}_{\refl}$ 
with the amplitude of the incident wave, $\vc{E}_{\inc}$
through the transmission and reflection
matrices, $\mvc{T}$ and $\mvc{R}$,
assume the following generalized form:
\begin{align}
  \label{eq:TR_gen}
    \begin{pmatrix}
\vc{E}_{\ind{out}}^{(-)}\\
\vc{E}_{\ind{out}}^{(+)}
\end{pmatrix}
=
\mvc{S}
\begin{pmatrix}
\vc{E}_{\ind{in}}^{(+)}\\
\vc{E}_{\ind{in}}^{(-)}
\end{pmatrix}
=
\begin{pmatrix}
\mvc{T}_{+} & \mvc{R}_{-}\\
\mvc{R}_{+} & \mvc{T}_{-}
\end{pmatrix}
\begin{pmatrix}
\vc{E}_{\ind{in}}^{(+)}\\
\vc{E}_{\ind{in}}^{(-)}
\end{pmatrix},
\end{align}
where 
$\mvc{S}$ is the matrix~---~ 
the so-called \textit{scattering matrix}~---~that relates 
the outgoing and incoming waves;
$\mvc{T}_{+}$ ($\mvc{R}_{+}$) is the transmission (reflection)
matrix for the case when the incident wave is incoming 
from the half space $z\le 0$
bounded by the entrance face,
whereas the mirror symmetric case where the incident wave is impinging onto 
the exit face of the sample is described by
the transmission (reflection) matrix $\mvc{T}_{-}$ ($\mvc{R}_{-}$).
So, we have
\begin{subequations}
 \begin{align}
&
 \label{eq:RT_pm}
\mvc{T}=\mvc{T}_{\pm},\quad
\mvc{R}=\mvc{R}_{\pm},
\\
   &
   \label{eq:inc}
\vc{E}_{\ind{in}}^{(\pm)}=
\vc{E}_{\inc}=
\begin{pmatrix}
E_{p}^{(\inc)}\\
E_{s}^{(\inc)}
\end{pmatrix},
\quad
\vc{E}_{\ind{in}}^{(\mp)}=0,
\\
&
   \label{eq:refl-transm}
   \vc{E}_{\ind{out}}^{(\pm)}=
\vc{E}_{\refl}\equiv
  \begin{pmatrix}
E_{p}^{(\refl)}\\
E_{s}^{(\refl)}
\end{pmatrix},
\quad
\vc{E}_{\ind{out}}^{(\mp)}=
\vc{E}_{\transm}\equiv
  \begin{pmatrix}
E_{p}^{(\transm)}\\
E_{s}^{(\transm)}
\end{pmatrix}.
 \end{align} 
\end{subequations}

It is our task now to relate
these matrices and the evolution operator
given by Eq.~\eqref{eq:evol_problem}.
To this end, we 
use  the boundary conditions requiring
the tangential components of the electric and magnetic
fields to be continuous at the boundary surfaces:
$\vc{F}(0)=\vc{F}_{\med}(0-0)$ and
$\vc{F}(h)=\vc{F}_{\med}(h+0)$,
and apply the relation~\eqref{eq:evol_problem}
to  the anisotropic layer of the thickness $D$
to yield the following result
\begin{align}
  \label{eq:continuity}
  \vc{F}_{\med}(h+0)=\mvc{U}(h,0)\cdot\vc{F}_{\med}(0-0),
\quad  
h=k_{\vac} D.
\end{align}

%%%%%%%%%%%%%%%%%%%%
\subsection{Transfer matrix}
\label{subsec:transf-matr}
%%%%%%%%%%%%%%%%%%%%

On substituting Eqs.~\eqref{eq:F_med}
into Eq.~\eqref{eq:continuity}
we have
\begin{align}
  \label{eq:transf-rel}
  \begin{pmatrix}
    \vc{E}_{\ind{in}}^{(+)}\\
    \vc{E}_{\ind{out}}^{(+)}
  \end{pmatrix}
=
\mvc{W}\cdot
  \begin{pmatrix}
    \vc{E}_{\ind{out}}^{(-)}\\
    \vc{E}_{\ind{in}}^{(-)}
  \end{pmatrix}
\end{align}
where the matrix $\mvc{W}$
linking the electric field vector amplitudes of the waves
in the half spaces $z<0$ and $z>D$
bounded by the faces of the layer
will be referred to as the \textit{transfer (linking) matrix}.
The expression for the transfer matrix is
as follows
\begin{align}
&
  \label{eq:W-op}
  \mvc{W}=
\mvc{V}_{\med}^{-1}\cdot\mvc{U}_{R}^{-1}(h)\cdot\mvc{V}_{\med}=
\begin{pmatrix}
\mvc{W}_{11} & \mvc{W}_{12}\\
\mvc{W}_{21} & \mvc{W}_{22}
\end{pmatrix}
\end{align}
where $\mvc{U}_{R}(\tau)=\mvc{T}_{\ind{rot}}(-\phi_p) \mvc{U}(\tau,0)
\mvc{T}_{\ind{rot}}(\phi_p)$
is the rotated operator of evolution.
This operator is the solution of the initial value
problem~\eqref{eq:evol_problem} 
with $\mvc{M}(\tau)$ replaced with
$\mvc{M}_R(\tau)=\mvc{T}_{\ind{rot}}(-\phi_p) \mvc{M}(\tau) \mvc{T}_{\ind{rot}}(\phi_p)$.

From Eqs.~\eqref{eq:TR_gen} and~\eqref{eq:transf-rel}, 
the block structure of the transfer matrix
can be expressed in terms of the transmission and reflection matrices
as follows  
\begin{align}
&  
\label{eq:W_TR}
\mvc{W}_{11}= \mvc{T}_{+}^{-1},
\quad
\mvc{W}_{12}=-\mvc{T}_{+}^{-1}\cdot\mvc{R}_{-},
\notag
\\
&
\mvc{W}_{21}=\mvc{R}_{+}\cdot\mvc{T}_{+}^{-1},
\quad
\mvc{W}_{22}=
\mvc{T}_{-}-\mvc{R}_{+}\cdot\mvc{T}_{+}^{-1}\cdot\mvc{R}_{-}.
\end{align}
Similarly, for inverse of the transfer matrix,
\begin{align}
  \label{eq:inv_W}
  \mvc{W}^{-1}=
\begin{pmatrix}
\mvc{W}_{11}^{(-1)}& \mvc{W}_{12}^{(-1)} \\
\mvc{W}_{21}^{(-1)} & \mvc{W}_{22}^{(-1)}
\end{pmatrix},
\end{align}
we have
\begin{align}
&
  \label{eq:inv_W_TR}
\mvc{W}_{11}^{(-1)}
=
\mvc{T}_{+}-\mvc{R}_{-}\cdot\mvc{T}_{-}^{-1}\cdot\mvc{R}_{+},
\quad
\mvc{W}_{12}^{(-1)} 
=
\mvc{R}_{-}\cdot\mvc{T}_{-}^{-1},
\notag
\\
&
\mvc{W}_{21}^{(-1)} 
=
-\mvc{T}_{-}^{-1}\cdot\mvc{R}_{+},
\quad
\mvc{W}_{22}^{(-1)}
=
\mvc{T}_{-}^{-1}.
\end{align}

%%%%%%%%%%%%%%%%%%%%%%
\subsection{Symmetries}
\label{subsec:symmetry}
%%%%%%%%%%%%%%%%%%%%%

In Appendix~\ref{sec:op-evol},
it is shown that,
for non-absorbing media with symmetric dielectric tensor, 
$\epsilon_{ij}=\epsilon_{ji}$,
the operator of evolution satisfies
the unitarity relation~\eqref{eq:unit_U}.
By using Eq.~\eqref{eq:unit_U}
in combination with
the algebraic identity
\begin{subequations}
\label{eq:Vm_identity}
\begin{align}
&  
\label{eq:idnt_Vm}
\tcnj{\mvc{V}}_{\med}
\cdot
\mvc{G}\cdot\mvc{V}_{\med}=
N_{\med}\mvc{G}_3,
\\
&
\label{eq:G_G3}
\mvc{G}=
\begin{pmatrix}
  \mvc{0}&\mvc{I}_2\\
\mvc{I}_2&\mvc{0}
\end{pmatrix},
\quad
\mvc{G}_3=\diag(\mvc{I}_2,-\mvc{I}_2),
\end{align}
\end{subequations}
where $N_{\med} = 2 q_{\med}/\mu_{\med}$,
for the eigenvector matrix given in Eq.~\eqref{eq:Vm-phi-q},
we can deduce
the unitarity relation for the transfer matrix~\eqref{eq:W-op}
\begin{align}
  \label{eq:unit_W}
  \mvc{W}^{-1}
=
\mvc{G}_3\cdot
\hcnj{\mvc{W}}
\cdot
\mvc{G}_3
=
\begin{pmatrix}
\hcnj{\mvc{W}}_{11} & -\hcnj{\mvc{W}}_{21}\\
-\hcnj{\mvc{W}}_{12} & \hcnj{\mvc{W}}_{22}
\end{pmatrix}.
\end{align}

The unitarity relation~\eqref{eq:unit_W} 
for non-absorbing layers
can now be used
to derive 
the energy conservation laws
\begin{subequations}
  \label{eq:energy_laws}
\begin{align}
&
  \label{eq:energy_consv}
  \hcnj{\mvc{T}}_{\pm}\mvc{T}_{\pm}+ 
\hcnj{\mvc{R}}_{\pm}\mvc{R}_{\pm}=\mvc{I}_2,
\\
&
  \label{eq:energy_pm}
  \mvc{T}_{\pm} \hcnj{\mvc{T}}_{\pm}+
 \mvc{R}_{\mp}\hcnj{\mvc{R}}_{\mp}
=
  \hcnj{[\tcnj{\mvc{T}}_{\pm}]} \tcnj{\mvc{T}}_{\pm}+
\hcnj{[\tcnj{\mvc{R}}_{\mp}]} \tcnj{\mvc{R}}_{\mp}
=\mvc{I}_2,
% \hcnj{\mvc{T}}_{+}\cdot\mvc{T}_{+}=
%   \hcnj{\mvc{T}}_{-}\cdot\mvc{T}_{-},
% \quad
% \hcnj{\mvc{R}}_{+}\cdot\mvc{R}_{+}=
% \hcnj{\mvc{R}}_{-}\cdot\mvc{R}_{-},
\end{align}
\end{subequations}
where a dagger and
the superscript $T$ 
will denote Hermitian conjugation
and matrix transposition, respectively,
along with  the relations for the block matrices 
\begin{subequations}
\label{eq:W_ij_unit}
\begin{align}
&
  \label{eq:W_ii_unit}
  \mvc{W}_{11}=\mvc{T}_{+}^{-1},\quad
  \mvc{W}_{22}=\hcnj{[\mvc{T}_{-}^{-1}]},
\\
&
 \label{eq:W_12_unit}
  \mvc{W}_{12}=-\mvc{T}_{+}^{-1} \mvc{R}_{-}=
\hcnj{[\mvc{T}_{-}^{-1} \mvc{R}_{+}]},
\\
&
 \label{eq:W_21_unit}
  \mvc{W}_{21}=\mvc{R}_{+} \mvc{T}_{+}^{-1}=
-\hcnj{[\mvc{R}_{-} \mvc{T}_{-}^{-1}]}.
\end{align}
\end{subequations}
Note that Eqs.~\eqref{eq:W_12_unit} and~\eqref{eq:W_21_unit}
can be conveniently rewritten in the following form
\begin{subequations}
\label{eq:rel-W_12_21}
\begin{align}
&
 \label{eq:W_12_unit2}
\mvc{T}_{-} \hcnj{\mvc{R}}_{-}= 
  -\mvc{R}_{+} \hcnj{\mvc{T}}_{+},
\\
&
 \label{eq:W_21_unit2}
\mvc{R}_{-} \mvc{T}_{-}^{-1}= 
-\hcnj{[\mvc{T}_{+}^{-1}]} \hcnj{\mvc{R}}_{+},
\end{align}
\end{subequations}
so that multiplying these identities and using the energy conservation
law~\eqref{eq:energy_consv} gives the relations~\eqref{eq:energy_pm}. 

In the translation invariant case of uniform anisotropy, 
the matrix $\mvc{M}$
is independent of $\tau$ and
the operator of evolution is given by
\begin{align}
  \label{eq:U-hom}
  \mvc{U}(\tau,\tau_0)=
\mvc{U}(\tau-\tau_0)=
     \exp\{i \mvc{M}\, (\tau-\tau_0)\}.
\end{align} 
Then, the unitarity condition~\cite{Kiselev:pra:2008}
\begin{align}
  \label{eq:unit_UW-hom}
  \mvc{U}^{-1}=\cnj{\mvc{U}},
\quad
\mvc{W}^{-1}=\cnj{\mvc{W}}
\end{align}
can be combined with Eq.~\eqref{eq:unit_W} 
to yield the additional symmetry relations
for $\mvc{W}_{ij}$
\begin{align}
  \label{eq:W_ij_hom}
  \tcnj{\mvc{W}}_{ii}=\mvc{W}_{ii},
\quad
  \tcnj{\mvc{W}}_{12}=-\mvc{W}_{21},
\end{align}
where an asterisk will indicate complex conjugation,
that give the following algebraic identities for the transmission and
reflection matrices:
\begin{align}
&
  \label{eq:TR-sym-hom}
   \tcnj{\mvc{T}}_{\pm}=\mvc{T}_{\pm},
\quad
   \tcnj{\mvc{R}}_{+}=\mvc{R}_{-},
\\
&
  \label{eq:T-sym-hom}
  \cnj{\mvc{T}}_{\pm}=-\cnj{\mvc{R}}_{\mp}
 \mvc{T}_{\mp} \mvc{R}_{\mp}^{-1}.
\end{align}
It can be readily seen that
the relation for the transposed matrices~\eqref{eq:energy_pm}
can be derived by
substituting Eq.~\eqref{eq:TR-sym-hom} 
into the conservation law~\eqref{eq:energy_consv}.

For the important special case of uniformly anisotropic planar structures
with $\mvc{M}_{11}=\mvc{M}_{22}=0$,
the algebraic structure of the transfer matrix 
is described in Appendix~\ref{sec:planar}.
Equation~\eqref{eq:tW_ij-pln}
shows that 
the symmetry relations~\eqref{eq:W_ij_hom}
remain valid even if
the dielectric constants are complex-valued
and 
the medium is absorbing.
Since identities~\eqref{eq:TR-sym-hom}
are derived from Eqs.~\eqref{eq:W_ij_hom}
and~\eqref{eq:W_TR} without recourse  to
the unitarity relations, they 
also hold for lossy materials.

Similar remark applies to
the expression for inverse of 
the transfer matrix~\eqref{eq:W_inv_pln}.
From Eq.~\eqref{eq:W_inv_pln}, 
it follows that 
the relation between
the transmission (reflection) matrix,
$\mvc{T}_{+}$ ($\mvc{R}_{+}$),
and its mirror symmetric counterpart
 $\mvc{T}_{-}$ ($\mvc{R}_{-}$)
can be further simplified and is given by
\begin{align}
  \label{eq:TR_pm_pln}
  \mvc{T}_{+}=
\bs{\sigma}_3 \mvc{T}_{-} \bs{\sigma}_3,
\quad
  \mvc{R}_{+}=
\bs{\sigma}_3 \mvc{R}_{-} \bs{\sigma}_3.
\end{align}
From Eqs.~\eqref{eq:TR_pm_pln} and~\eqref{eq:TR-sym-hom},
we have the relation
for the transposed reflection matrices
\begin{align}
  \label{eq:R_transp-pln}
   \tcnj{\mvc{R}}_{\pm}=\bs{\sigma}_3 \mvc{R}_{\pm} \bs{\sigma}_3,
\end{align}
whereas the transmission matrices are symmetric. 

%%%%%%%%%%%%%%%%%%%%%%%%%%%%
\section{Electro-optics of homogenized DHFLC cells}
\label{sec:optics-dhf}
%%%%%%%%%%%%%%%%%%%%%%%%%%%

We now pass on to the electro-optical properties of DHFLC cells
and extend the results of Ref.~\cite{Kiselev:pre:2011}
to the case of biaxial ferroelectric liquid crystals with
subwavelength pitch.
In addition,
the theoretical treatment will be significantly improved
by using an alternative fully consistent procedure to perform 
averaging over distorted FLC helix that goes around
the limitations of the first-order approximation.

%%%%%%%%%%%%%%%%%%%%%%%%%%%%
\subsection{Effective dielectric tensor}
\label{subsec:diel-tensor-dhf}
%%%%%%%%%%%%%%%%%%%%%%%%%%%

We consider a FLC film of thickness $D$
with the $z$ axis 
which, as is indicated in Fig.~\ref{fig:geom}, 
is normal to the bounding surfaces: $z=0$ and $z=D$,
and introduce
the effective  dielectric tensor,
$\bs{\varepsilon}_{\eff}$,
describing a homogenized DHFLC helical structure.
For a biaxial FLC, the components of the dielectric tensor,
$\bs{\varepsilon}$,
are given by
\begin{align}
&
\epsilon_{i j}=
  \epsilon_{\perp}
\delta_{i j}+
(\epsilon_{1}-\epsilon_{\perp})\,
d_i d_j
+
(\epsilon_{2}-\epsilon_{\perp})\,
p_i p_j
\notag
\\
&
=
  \epsilon_{\perp}
(
\delta_{i j}+u_1 d_i d_j
+u_2 p_i p_j
),
\label{eq:diel-tensor}
\end{align}
where 
$i,j\in\{x,y,z\}$,
$\delta_{ij}$ is the Kronecker symbol;
$d_i$ ($p_i$) is the $i$th component
of the FLC director (unit polarization vector)
given by Eq.~\eqref{eq:director} (Eq.~\eqref{eq:pol-vector});
$u_i=(\epsilon_{i}-\epsilon_{\perp})/\epsilon_{\perp}=\Delta\epsilon_i/\epsilon_{\perp}=r_i-1$
are the \textit{anisotropy parameters}
and $r_1=\epsilon_1/\epsilon_\perp$ ($r_2=\epsilon_2/\epsilon_\perp$)
is the \textit{anisotropy (biaxiality) ratio}.
Note that, in the case of uniaxial anisotropy with $u_2=0$,
the principal values of the dielectric tensor are:
$\epsilon_2=\epsilon_{\perp}$ and
$\epsilon_{1}=\epsilon_{\parallel}$,
where $n_{\perp}=\sqrt{\mu\epsilon_{\perp}}$
($n_{\parallel}=\sqrt{\mu\epsilon_{\parallel}}$)
is the ordinary (extraordinary) refractive index
and the magnetic tensor of FLC 
is assumed to be isotropic with the magnetic permittivity $\mu$. 
As in Sec.~\ref{sec:theory} (see Fig.~\ref{fig:geom_4rays}), 
the medium surrounding the layer is optically
isotropic and is characterized by 
the dielectric constant $\epsilon_{\med}$,
the magnetic permittivity $\mu_{\med}$ and the refractive index
$n_{\med}=\sqrt{\mu_{\med} \epsilon_{\med}}$.

At $E=0$, the ideal FLC helix 
\begin{align}
&
\Phi= q_0 x\equiv \phi_0,
\label{eq:Phi_0}    
 \end{align}
where $q_0=2\pi/P_0$ is the free twist wave number
and $P_0$ is the equilibrium helical pitch,
is defined through the azimuthal angle around the smectic cone
$\Phi$ (see Fig.~\ref{fig:geom} and Eq.~\eqref{eq:director-planar}) 
and represents the undistorted structure.
For sufficiently small electric fields,
the standard perturbative technique
applied to the Euler-Lagrange equation
gives the first-order expression~\cite{Chigr:1999,Hedge:lc:2008}
for the azimuthal angle of a weakly distorted helical structure
\begin{align}
&
\Phi=\phi_0-\beta_E\sin\phi_0,
\label{eq:Phi_weak}    
 \end{align}
where $\beta_E=\gamma_E E$ is 
the electric field parameter linearly proportional to
the ratio of the applied and critical electric
fields: $E/E_c$, and $P=P_0$.

 According to Ref.~\cite{Kiselev:pre:2011},
normally incident light feels 
effective in-plane anisotropy
described by the averaged tensor,
$\avr{\bs{\varepsilon}_P}$:
\begin{align}
&
\avr{\epsilon_{\alpha\beta}^{(P)}}=
\left\langle
 \epsilon_{\alpha\,\beta}-\frac{ \epsilon_{\alpha\,z}
   \epsilon_{z\,\beta}}{ \epsilon_{zz}}
\right\rangle
\notag
\\
&
=
\epsilon_{0}
\left\langle
\delta_{\alpha\beta}+\frac{u_1 d_\alpha d_\beta+u_2 p_\alpha
  p_\beta+u_1 u_2 q_\alpha q_\beta}{1+u_1  d_z^2+u_2  p_z^2}\,
\right\rangle,
\label{eq:in-plane}
\\
&
q_{\alpha}=p_z d_{\alpha}-d_z p_{\alpha},
\quad 
\alpha,\beta\in\{x,y\},
\label{eq:q-vector}
\end{align}
where $\avr{\ldots}\equiv\avr{\ldots}_{\phi}=(2\pi)^{-1}\int_{0}^{2\pi}\ldots\dd\phi$,
and the effective dielectric tensor
\begin{align}
  \label{eq:eff-diel-tensor}
&
\bs{\varepsilon}_{\eff}=
\begin{pmatrix}
  \epsilon_{xx}^{(\eff)} & \epsilon_{xy}^{(\eff)}& \epsilon_{xz}^{(\eff)}\\
\epsilon_{yx}^{(\eff)}& \epsilon_{yy}^{(\eff)} & \epsilon_{yz}^{(\eff)}\\
\epsilon_{zx}^{(\eff)} & \epsilon_{zy}^{(\eff)} & \epsilon_{zz}^{(\eff)}
\end{pmatrix}  
\end{align}
can be expressed in terms of the averages
    \begin{align}
&
   \eta_{zz}=
\avr{\epsilon_{zz}^{-1}}
=
\epsilon_{0}^{-1}
\avr{[1+u_1 d_z^2+u_2 p_z^2]^{-1}},
\label{eq:eta}
\\
&
    \beta_{z\alpha}=
\avr{\epsilon_{z\alpha}/\epsilon_{zz}}
=
\left\langle
\frac{u_1 d_z d_\alpha+u_2 p_z p_\alpha}{1+u_1 d_z^2+u_2 p_z^2}
\right\rangle,
\label{eq:beta}
  \end{align}
as follows
\begin{align}
&
\epsilon_{zz}^{(\eff)}
=1/\eta_{zz},
\quad
\epsilon_{z\alpha}^{(\eff)}
=\beta_{z\alpha}/\eta_{zz},
\notag
\\
&
\epsilon_{\alpha\beta}^{(\eff)}
=\avr{\epsilon_{\alpha\beta}^{(P)}}+
\beta_{z\alpha} \beta_{z\beta}/\eta_{zz}.
\label{eq:elements-eff-diel-tensor}
  \end{align}
 
General
formulas~\eqref{eq:in-plane}-\eqref{eq:elements-eff-diel-tensor}
give  the zero-order approximation
for homogeneous models
describing the optical properties of
short pitch DHFLCs~\cite{Kiselev:pre:2011,Kiselev:pre:2013}.
Assuming that
the pitch-to-wavelength ratio $P/\lambda$
is sufficiently small,
these formulas 
can now be used to derive the effective dielectric tensor of
homogenized short-pitch DHFLC cell for both vertically and planar aligned
FLC helix.
The results for vertically aligned DHFLC cells
were recently published in Ref.~\cite{Kiselev:pre:2013}
and we concentrate on the geometry
of planar aligned DHFLC helix
shown in Fig.~\ref{fig:geom}.
For this geometry, the parameters needed to compute 
the averages $\avr{\bs{\varepsilon}_P}$
(see Eq.~\eqref{eq:in-plane}),
$\avr{\eta_{zz}}$ 
(see Eq.~\eqref{eq:eta})
and $\avr{\beta_{z\alpha}}$
(see Eq.~\eqref{eq:beta})
are given by
\begin{align}
  \label{eq:d_i-planar}
&
d_z=\sin\theta\sin\Phi,
\quad
\begin{pmatrix}
  d_x\\
d_y
\end{pmatrix}
=
\begin{pmatrix}
  \cos\theta\\
\sin\theta\cos\Phi
\end{pmatrix}
\\
&
\label{eq:p_i-planar}
p_z=\cos\Phi,
\quad
\begin{pmatrix}
  p_x\\
p_y
\end{pmatrix}
=
\begin{pmatrix}
  0\\
-\sin\Phi
\end{pmatrix}
,
\quad
\begin{pmatrix}
  q_x\\
q_y
\end{pmatrix}
=
\begin{pmatrix}
  \cos\theta\cos\Phi\\
  \sin\theta
\end{pmatrix}
\\
&
\label{eq:e_zz-planar}
  \frac{\epsilon_{zz}}{\epsilon_{2}}\equiv
v_{zz}=1+v \sin^2\Phi,\quad
v=v_1 \sin^2\theta -v_2,
\quad
v_i=u_i/r_2=\Delta\epsilon_i/\epsilon_2.
\end{align}

Formulas~\eqref{eq:epsilon-ij-planar}
can now be inserted
into Eqs.~\eqref{eq:elements-eff-diel-tensor}
to yield the explicit expressions for the elements
of the dielectric tensor~\eqref{eq:eff-diel-tensor}: 
\begin{subequations}
  \label{eq:epsilon-ij-planar}
\begin{align}
&
\label{eq:epsilon_zz-xz-planar}
\epsilon_{z z}^{(\eff)}
=\epsilon_2/\avr{v_{zz}^{-1}},
\quad
 \begin{pmatrix} 
\epsilon_{z x}^{(\eff)}
/\epsilon_{z z}^{(\eff)}
\\
\epsilon_{z y}^{(\eff)}
/\epsilon_{z z}^{(\eff)}
\end{pmatrix}
=
\begin{pmatrix}
v_1 \cos\theta\sin\theta  \avr{v_{zz}^{-1}\sin\Phi}\\
  v \avr{v_{zz}^{-1}\sin\Phi\cos\Phi}
\end{pmatrix}, 
\\
  \label{eq:epsilon_xx-planar}
  &
    \epsilon_{xx}^{(\eff)}/\epsilon_{\perp}
=1+(r_1/r_2-1-v)
\avr{v_{zz}^{-1}(1+u_2\cos^2\Phi)},
\\
&
\label{eq:epsilon_yy-planar}
    \epsilon_{yy}^{(\eff)}/\epsilon_{\perp}=
1+v\avr{v_{zz}^{-1}\cos^2\Phi} +
u_2(1+v)\avr{v_{zz}^{-1}},
\\
&
\label{eq:epsilon_xy-planar}
     \epsilon_{xy}^{(\eff)}/\epsilon_{\perp}=
u_1\cos\theta\sin\theta
\avr{v_{zz}^{-1}\cos\Phi}.
\end{align}
\end{subequations}

%%%%%%%%%%%%%%%%%%%%%%
\subsection{Orientational Kerr  effect}
\label{subsec:kerr-effect}
%%%%%%%%%%%%%%%%%%%%%

The simplest averaging procedure 
previously used in Refs.~\cite{Abdul:mclc:1991,Kiselev:pre:2011,Kiselev:pre:2013}
involves substituting
the formula for a weakly distorted FLC helix~\eqref{eq:Phi_weak}
into Eqs.~\eqref{eq:epsilon-ij-planar} and performing integrals
over $\phi_0$. 
This procedure thus
heavily relies on the first-order approximation where the director
distortions are described by the term linearly proportional to the
electric field (the second term on the right hand side of Eq.~\eqref{eq:Phi_weak}).
Quantitatively, the difficulty with this approach is that
the linear approximation may not be suffice
for accurate computing of the second-order contributions
to the diagonal elements of the dielectric
tensor~\eqref{eq:epsilon-ij-planar}.
In this approximation, the
second-order corrections describing
the helix distortions that involve the change of the helix pitch
have been neglected. 

In order to circumvent the problem, in this paper,
we apply
an alternative approach that allows to go beyond
the first-order approximation without recourse to
explicit formulas for the azimuthal angle.
This method is detailed in Appendix~\ref{sec:FLC-spiral}.
The analytical results~\eqref{eq:averages-planar}
substituted into Eqs.~\eqref{eq:epsilon-ij-planar}
give the effective dielectric tensor
in the following form:
\begin{align}
  \label{eq:eff-diel-planar}
  &
\bs{\varepsilon}_{\eff}=
\begin{pmatrix}
  \epsilon_h+\gamma_{xx}\alpha_E^2 & \gamma_{xy} \alpha_E& 0\\
  \gamma_{xy} \alpha_E,& \epsilon_p+\gamma_{yy}\alpha_E^2 & 0\\
0 & 0 & \epsilon_p-\gamma_{yy}\alpha_E^2
\end{pmatrix}.
\end{align}
The \textit{zero-field dielectric constants}, 
$\epsilon_h$ and $\epsilon_p$,
that enter the tensor~\eqref{eq:eff-diel-planar} are defined in 
Eqs.~\eqref{eq:epsilon_h-v} and~\eqref{eq:epsilon_p-v},
respectively,
and can be conveniently rewritten as follows
\begin{subequations}
  \label{eq:epsilon_ph}
\begin{align}
&
\label{eq:epsilon_h}
\epsilon_h
/\epsilon_{\perp}
=
r_2^{-1/2}
\biggl\{
\sqrt{r_2}+u_1 \cos^2\theta
\left(
\frac{r_2-1}{\sqrt{u}+\sqrt{r_2}}+u^{-1/2}
\right)
\biggr\},
\\
&
\label{eq:epsilon_p}
\epsilon_p/\epsilon_{\perp}=
\sqrt{r_2 u},
\quad
u= r_2 (v+1)=
u_1\sin^2\theta+1.
\end{align}
\end{subequations}
A similar result for 
the \textit{coupling coefficients}
$\gamma_{xx}$,
$\gamma_{yy}$ and $\gamma_{xy}$
(see Eq.~\eqref{eq:coupling-coeffs-v})
reads
\begin{subequations}
  \label{eq:coupling-coeffs}
   \begin{align}
&
\label{eq:gxx-u}
\gamma_{xx}/\epsilon_{\perp}=
\frac{3\sqrt{r_2/u}}{(\sqrt{u}+\sqrt{r_2})^2}
(u_1\cos\theta\sin\theta)^2,
%(r_1-v_1)(v_1-1),
\\
&
\label{eq:gyy-u}
\gamma_{yy}/\epsilon_{\perp}=
\frac{3\sqrt{r_2u}}{(\sqrt{u}+\sqrt{r_2})^2}
(u-r_2),
\\
&
\label{eq:gxy-u}
\gamma_{xy}/\epsilon_{\perp}=
\frac{2\sqrt{r_2}}{\sqrt{u}+\sqrt{r_2}}
u_1\cos\theta\sin\theta.
    \end{align}
\end{subequations}

Note that, following Ref.~\cite{Kiselev:pre:2013},
we have 
used the relation~\eqref{eq:avr-cos} to
introduce the electric field parameter
\begin{align}
  \label{eq:alpha_E}
\alpha_E=\chi_E E/P_s,  
\end{align}
where $\chi_E=\partial \avr{P_z}/\partial E$ is
the dielectric susceptibility of 
the Goldstone mode~\cite{Carlsson:pra:1990,Urbanc:ferro:1991}.

The above dielectric tensor
is characterized by the three generally different principal values
(eigenvalues)
and the corresponding optical axes (eigenvectors)
as follows
\begin{align}
&
  \label{eq:eff-diel-diag-planar}
  \bs{\varepsilon}_{\eff}=
\epsilon_z \uvc{z}\otimes\uvc{z}
+\epsilon_{+} \uvc{d}_{+}\otimes\uvc{d}_{+}
+\epsilon_{-} \uvc{d}_{-}\otimes\uvc{d}_{-},
\\
&
\label{eq:epsilon_z}
\epsilon_{z}=n_{z}^{\,2}=\epsilon_{zz}^{(\eff)}=\epsilon_p-\gamma_{yy}\alpha_E^2,
\\
&
\label{eq:epsilon_pm}
\epsilon_{\pm}=n_{\pm}^{\,2}=\bar{\epsilon}\pm\sqrt{[\Delta\epsilon]^2+[\gamma_{xy} \alpha_{E}]^2}
\end{align}
where
\begin{align}
&
  \label{eq:epsilon_avr}
  \bar{\epsilon}=(\epsilon_{xx}^{(\eff)}+\epsilon_{yy}^{(\eff)})/2
=\bar{\epsilon}_0+(\gamma_{xx}+\gamma_{yy})\alpha_{E}^2/2,
\quad
\bar{\epsilon}_0=(\epsilon_{h}+\epsilon_{p})/2,
\\
&
  \label{eq:Delta_epsilon}
 \Delta\epsilon=(\epsilon_{xx}^{(\eff)}-\epsilon_{yy}^{(\eff)})/2=
\Delta\epsilon_0+(\gamma_{xx}-\gamma_{yy})\alpha_{E}^2/2,
\quad
\Delta\epsilon_0=(\epsilon_{h}-\epsilon_{p})/2.
\end{align}
The in-plane optical axes are given by
\begin{align}
  \label{eq:d_plus}
  \uvc{d}_{+}=\cos\psi_d\,\uvc{x}+
\sin\psi_d\,\uvc{y},
\quad
  \uvc{d}_{-}=\uvc{z}\times  \uvc{d}_{+},
\quad
2\psi_d =\arg[\Delta\epsilon +i \gamma_{xy} \alpha_E].
\end{align}
From Eq.~\eqref{eq:eff-diel-planar}, it is clear that, 
similar to the case of uniaxial FLCs studied in Ref.~\cite{Kiselev:pre:2011},
the zero-field dielectric tensor is
uniaxially anisotropic with the optical axis directed along the
twisting axis $\uvc{h}=\uvc{x}$. The applied electric field
changes the principal values 
(see Eqs.~\eqref{eq:epsilon_z} and~\eqref{eq:epsilon_pm})
so that 
the electric-field-induced anisotropy is generally biaxial.
In addition, the in-plane principal optical axes are rotated about the
vector of electric field, $\vc{E}\parallel \uvc{z}$,
by the angle $\psi_d$ given in Eq.~\eqref{eq:d_plus}.

In the low electric field  region,
the electrically induced part of the principal values
is typically dominated by the
Kerr-like nonlinear terms proportional to $E^2$, whereas 
the electric field dependence of the angle $\psi_d$ is
approximately linear: $\psi_d\propto E$.
This effect is caused by
the electrically induced distortions of the helical
structure and bears some resemblance to the electro-optic Kerr effect. 
Following Refs.~\cite{Kiselev:pre:2013,Kiselev:ol:2014}, 
it will be referred to as the \textit{orientational Kerr effect}.

It should be emphasized that this effect differs from
the well-known Kerr effect which is  a quadratic electro-optic effect
related to the electrically induced birefringence
in optically isotropic (and transparent) materials
and
which is mainly caused by the electric-field-induced orientation
of polar molecules~\cite{Weinberger:pml:2008}.
By contrast, in our case, similar to 
polymer stabilized blue phase liquid crystals~\cite{Yan:apl:2010,Yan:apl:2013},
we deal with the effective dielectric tensor
of a nanostructured chiral smectic liquid crystal. 
This tensor~\eqref{eq:epsilon-ij-planar} is defined
through averaging over the FLC orientational structure.

\begin{figure*}[!tbh]
\centering
  %\resizebox{95mm}{!}{\includegraphics*{expt-th_130_comp.eps}}
  \resizebox{95mm}{!}{\includegraphics*{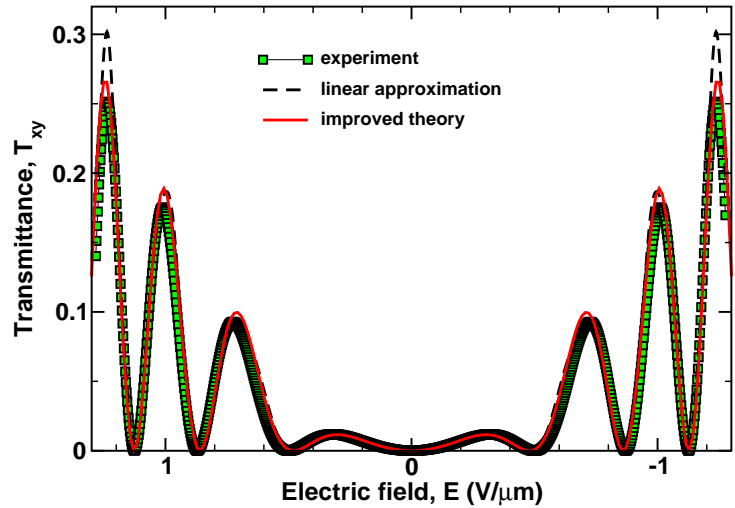}}
\caption{%
(Color online)
Transmittance of light passing through crossed polarizers, 
$T_{xy}$, as a function of the applied electric
field at the wavelength $\lambda=650$~nm 
for the DHFLC cell of thickness $D=130$~\mum\
filled with the FLC mixture FLC-576A~\cite{Kiselev:pre:2011}.
Parameters of the mixture are: 
$n_{\perp}=\sqrt{\epsilon_{\perp}}=1.5$ 
($n_{\parallel}=\sqrt{\epsilon_1}=1.72$) is the ordinary (extraordinary)
refractive index and $\theta=32$~deg is the tilt angle.
The experimental points are marked by squares.
The results obtained using the linear approximation
are shown as dashed line
Dashed and solid lines represent the theoretical curves
computed using the linear approximation~\cite{Kiselev:pre:2011} 
and the improved method of averaging with
$P_s/\chi_E\approx 3.4$~V/\mum, respectively.
}
\label{fig:expt_th}
\end{figure*}

Typically, in experiments
dealing with
the electro-optic response
of DHFLC cells,  
the transmittance of normally incident 
light passing through crossed polarizers
is measured as a function of the applied electric field.
For normal incidence,
the transmission and reflection matrices can be easily obtained
from the results given in Appendix~\ref{sec:planar} 
by substituting Eq.~\eqref{eq:Apm_norm_pln}
into Eqs.~\eqref{eq:tW_ij-pln}-~\eqref{eq:TR_pln}.
When the incident wave is linearly polarized along the
$x$ axis (the helix axis),
the transmittance coefficient 
\begin{align}
&
  \label{eq:T-xy}
  T_{xy}=|t_{xy}|^2=\frac{|t_{+}-t_{-}|^2}{4}\,\sin^2(2\psi_\dd),
\quad
\sin^2(2\psi_\dd)=\frac{\alpha_{E}^2}{\alpha_{E}^2+(\Delta\epsilon/\gamma_{xy})^2}\,,
\\
&
 \label{eq:t-pm}
  t_{\pm}=\frac{1-\rho_{\pm}^2}{%
1-\rho_{\pm}^2\exp(2in_{\pm}h)
}\exp(i n_{\pm} h),
\quad
\rho_{\pm}=\frac{n_{\pm}/\mu-n_{\med}/\mu_{\med}}{%
n_{\pm}/\mu+n_{\med}/\mu_{\med}}.
\end{align}
where $h=k_{\vac} D$ is the thickness parameter,
describes the intensity of the light passing through 
crossed polarizers.
Note that, under certain conditions such as $|\rho_{\pm}|\ll 1$,
$t_{\pm}\approx \exp(i n_{\pm} h)$ and 
the transmittance~\eqref{eq:T-xy} can be approximated by simpler formula
\begin{align}
  \label{eq:T-xy-approx}
 T_{xy} \approx\sin^2(\delta/2)\,\sin^2(2\psi_\dd), 
\end{align}
where
$\delta=\Delta n_{\eff}\, h=(n_{+}-n_{-}) h$
is the difference in optical path of the ordinary and extraordinary
waves known as the \textit{phase retardation}.

In Ref.~\cite{Kiselev:pre:2011},
the relation~\eqref{eq:T-xy} was used to 
fit  the experimental data using
the theory based on the linear approximation for 
the helix distortions (see Eq.~\eqref{eq:Phi_weak}).
These results are reproduced in Figure~\ref{fig:expt_th}
along with the theoretical curve computed 
using the modified averaging technique.
From Fig.~\ref{fig:expt_th},
it is seen that,
in the range of relatively high voltages,
the averaging method described in 
Appendix~\ref{sec:FLC-spiral} 
improves agreement between the theory and 
the experiment, whereas, at small voltages,
the difference between the fitting curves is negligibly small.

\begin{figure*}[!tbh]
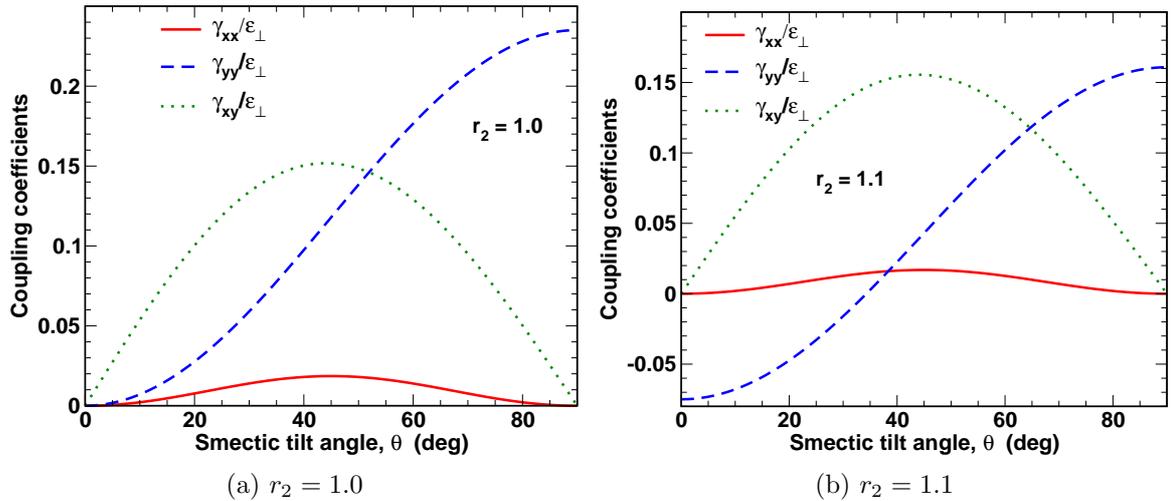

\centering
\subfloat[$r_2=1.0$]{
 % \resizebox{75mm}{!}{\includegraphics*{gamma_theta_uni.eps}}
  \resizebox{75mm}{!}{\includegraphics*{fig4a.eps}}
\label{subfig:gamma_uni}
}
\subfloat[$r_2=1.1$]{
%  \resizebox{75mm}{!}{\includegraphics*{gamma_theta_bia.eps}}
\resizebox{75mm}{!}{\includegraphics*{fig4b.eps}}
\label{subfig:gamma_bia}
}
\caption{%
(Color online)
Coupling coefficients
as a function of the smectic tilt angle $\theta$
at $r_1=(1.72/1.5)^2\approx 1.32$
for (a)~a uniaxially anisotropic FLC with $r_2=1$
and (b)~a biaxially anisotropic FLC with $r_2=1.1$. 
}
\label{fig:gamma_theta}
\end{figure*}

%%%%%%%%%%%%%%%%%%%%%%%%%%%%
\subsection{Effects of smectic tilt angle}
\label{subsec:tilt}
%%%%%%%%%%%%%%%%%%%%%%%%%%%

Given the anisotropy and biaxiality ratios, $r_1$ and $r_2$,
the zero-field dielectric constants ~\eqref{eq:epsilon_ph}
and the coupling coefficients~\eqref{eq:coupling-coeffs}
are determined by the smectic tilt angle, $\theta$.
Figure~\ref{fig:gamma_theta} shows how the coupling coefficients
depend on $\theta$ for both uniaxially and biaxially anisotropic
FLCs. 

As it can be seen in Fig.~\ref{subfig:gamma_uni}, 
in the case of conventional FLCs with $r_2=1$,
all the coefficients are positive
and
the difference of the coupling constants
$\gamma_{xx}-\gamma_{yy}$
that define the electrically induced part of $\Delta\epsilon$
(see Eq.~\eqref{eq:Delta_epsilon})
is negative at $0<\theta<\pi/2$.

From Eqs.~\eqref{eq:epsilon_z} and~\eqref{eq:epsilon_pm},
it follows that, at $r_2=1$, 
the principal values of dielectric constants
$\epsilon_z$ and $\epsilon_{-}$ are decreasing functions of
the electric field parameter $\alpha_E$ so that anisotropy of the effective
dielectric tensor~\eqref{eq:eff-diel-planar} 
is weakly biaxial.
In addition, for non-negative $\alpha_E$ and $\gamma_{xx}-\gamma_{yy}<0$,
the azimuthal angle of in-plane optical axis,
$\psi_{\dd}$, given in Eq.~\eqref{eq:d_plus} increases with $\alpha_E$
from zero to $\pi/2$.

Figure~\ref{subfig:gamma_bia} demonstrates that this is no longer 
the case for biaxial FLCs. It is seen that, at $r_2=1.1$,
the coupling coefficient $\gamma_{yy}$ and the difference
$\gamma_{xx}-\gamma_{yy}$ both change in sign
when  the tilt angle $\theta$ is sufficiently small.
At such angles, the dielectric constant $\epsilon_z$ increases 
with $\alpha_E$ and the electric field induced 
anisotropy of DHFLC cell becomes strongly biaxial.
When $\gamma_{xx}-\gamma_{yy}$ and $\alpha_E$ are positive,
electric field dependence of the azimuthal angle $\psi_{\dd}$
is non-monotonic and the angle decays to zero in the range of 
high voltages where $\psi_{\dd}\propto \alpha_{E}^{-1}$. 

\begin{figure*}[!tbh]
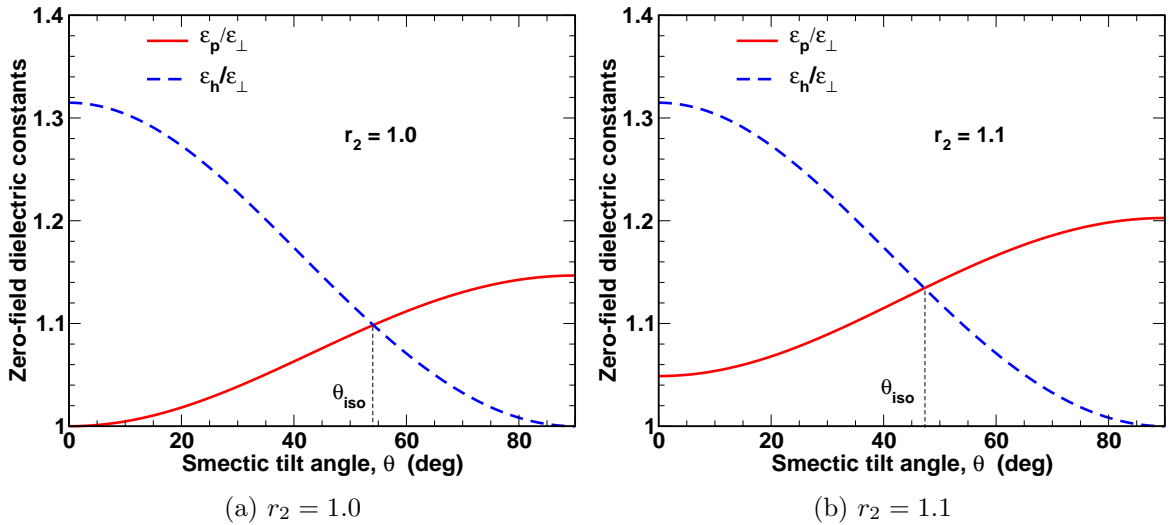

\centering
\subfloat[$r_2=1.0$]{
  %\resizebox{75mm}{!}{\includegraphics*{eph_theta_uni.eps}}
  \resizebox{75mm}{!}{\includegraphics*{fig5a.eps}}
\label{subfig:eph_uni}
}
\subfloat[$r_2=1.1$]{
  %\resizebox{75mm}{!}{\includegraphics*{eph_theta_bia.eps}}
\resizebox{75mm}{!}{\includegraphics*{fig5b.eps}}
\label{subfig:eph_bia}
}
\caption{%
(Color online)
Zero-field dielectric constants
as a function of the smectic tilt angle $\theta$
at $r_1=(1.72/1.5)^2\approx 1.32$
for (a)~a uniaxially anisotropic FLC with $r_2=1$
and (b)~a biaxially anisotropic FLC with $r_2=1.1$. 
}
\label{fig:eph_theta}
\end{figure*}

%%%%%%%%%%%%%%%%%%%%%%%%%%%%
\subsection{Zero-field isotropy and electro-optic response near
  exceptional point}
\label{subsec:isotropy}
%%%%%%%%%%%%%%%%%%%%%%%%%%%

At $E=0$, the zero-field anisotropy is uniaxial and is 
described by the dielectric constants, $\epsilon_h$ and
$\epsilon_{p}$, given in Eq.~\eqref{eq:epsilon_ph}. 
In Fig.~\ref{fig:eph_theta}, 
these constants are plotted against the tilt angle.
It is shown that, at small tilt angles, 
the anisotropy $\epsilon_{h}-\epsilon_{p}$ is positive.
It decreases with $\theta$ and
the zero-field state becomes isotropic
when,
at certain critical angle $\theta=\theta_{\ind{iso}}$,
the condition of zero-field isotropy
\begin{align}
  \label{eq:isotropy-1}
  \epsilon_p=\epsilon_z
\end{align}
is fulfilled 
and $\Delta\epsilon_0=0$.
So, the angle $\theta_{\ind{iso}}$
can be referred to as the \textit{isotropization angle}.
In what follows we discuss 
peculiarities of
the electro-optic response in the vicinity
of the \textit{isotropization point} where
$\Delta\epsilon$ is proportional to $E^2$ 
(see Eq.~\eqref{eq:Delta_epsilon})
and the Kerr-like regime breaks down.  

\begin{figure*}[!tbh]
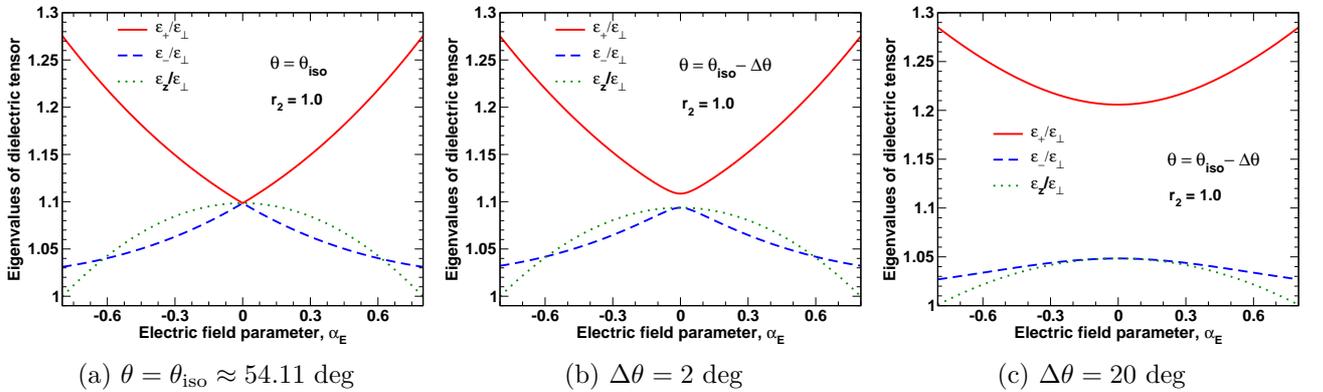

\centering
\subfloat[$\theta=\theta_{\ind{iso}}\approx 54.11$~deg]{
  %\resizebox{55mm}{!}{\includegraphics*{epsilon_alp_c_uni.eps}}
  \resizebox{55mm}{!}{\includegraphics*{fig6a.eps}}
\label{subfig:epsilon_c_uni}
}
\subfloat[$\Delta\theta=2$~deg]{
  %\resizebox{55mm}{!}{\includegraphics*{epsilon_alp_c-2_uni.eps}}
  \resizebox{55mm}{!}{\includegraphics*{fig6b.eps}}
\label{subfig:epsilon_c-2_uni}
}
\subfloat[$\Delta\theta=20$~deg]{
  %\resizebox{55mm}{!}{\includegraphics*{epsilon_alp_c-20_uni.eps}}
  \resizebox{55mm}{!}{\includegraphics*{fig6c.eps}}
\label{subfig:epsilon_c-20_uni}
}
\caption{%
(Color online)
Principal values of the effective dielectric tensor
as a function of the electric field parameter at 
 $r_1\approx 1.32$ and $r_2=1$
for different values of the smectic tilt angles. 
}
\label{fig:epsilon_alpha_uni}
\end{figure*}

Mathematically, the isotropization point represents a square root branch-point singularity
of the eigenvalues~\eqref{eq:epsilon_pm} of the dielectric tensor
which is known as 
the \textit{exceptional point}~\cite{Kato:bk:1995,Heiss:jpa:1990,Heiss:pre:2000}.
In the electric field dependence of the in-plane dielectric constants,
$\epsilon_{+}$ and $\epsilon_{-}$,
this singularity reveals itself as a cusp
where the derivatives of $\epsilon_{\pm}$ with respect to $\alpha_E$ are discontinuous.
More precisely, we have
\begin{align}
&
\label{eq:cusp}
\pdr{\epsilon_{+}}{\alpha_E}\biggl\lvert_{\alpha_E=0\pm 0}=
-\pdr{\epsilon_{-}}{\alpha_E}\biggl\lvert_{\alpha_E=0\pm 0}=
\pm |\gamma_{xy}|.      
    \end{align}
As is illustrated in Fig.~\ref{subfig:epsilon_c_uni}, 
the cusp is related to the effect of reconnection
of different branches representing solutions of 
an algebraic equation.

\begin{figure*}[!tbh]
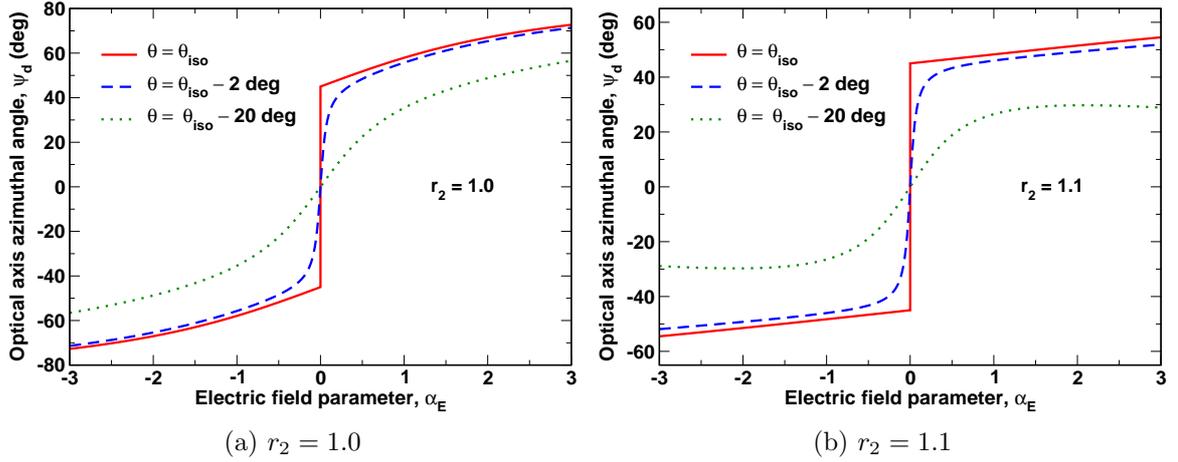

\centering
\subfloat[$r_2=1.0$]{
  %\resizebox{75mm}{!}{\includegraphics*{phi_alp_uni.eps}}
\resizebox{75mm}{!}{\includegraphics*{fig7a.eps}}
\label{subfig:psi_uni}
}
\subfloat[$r_2=1.1$]{
  %\resizebox{75mm}{!}{\includegraphics*{phi_alp_bia.eps}}
\resizebox{75mm}{!}{\includegraphics*{fig7b.eps}}
\label{subfig:psi_bia}
}
\caption{%
(Color online)
Principal axis azimuthal angle
versus the electric field parameter
at $r_1\approx 1.32$
for (a)~a uniaxially anisotropic FLC with $r_2=1$
and (b)~a biaxially anisotropic FLC with $r_2=1.1$. 
}
\label{fig:psi_alp}
\end{figure*}

Since the azimuthal angle $\psi_{\dd}$ is undetermined at $\theta=\theta_{\ind{iso}}$
\begin{align}
&
\label{eq:phase_sing}
      (\Delta\epsilon+i\gamma_{xy}\alpha_E)\Bigl\lvert_{\alpha_E=0}=0
\Longrightarrow \psi_d - ?
    \end{align}
the isotropization point also represents a phase singularity.
The electric field dependence of $\psi_{\dd}$ is thus discontinuous
and the relation 
    \begin{align}
&
\label{eq:phase_jump}
\psi_d\Bigl\lvert_{\alpha_E=0+0}-\psi_d\Bigl\lvert_{\alpha_E=0-0}=\sign(\gamma_{xy})\,\frac{\pi}{2}      
    \end{align}
describes its jumplike behaviour at $E=0$.
This behaviour is demonstrated in Fig.~\ref{fig:psi_alp}.

\begin{figure*}[!tbh]
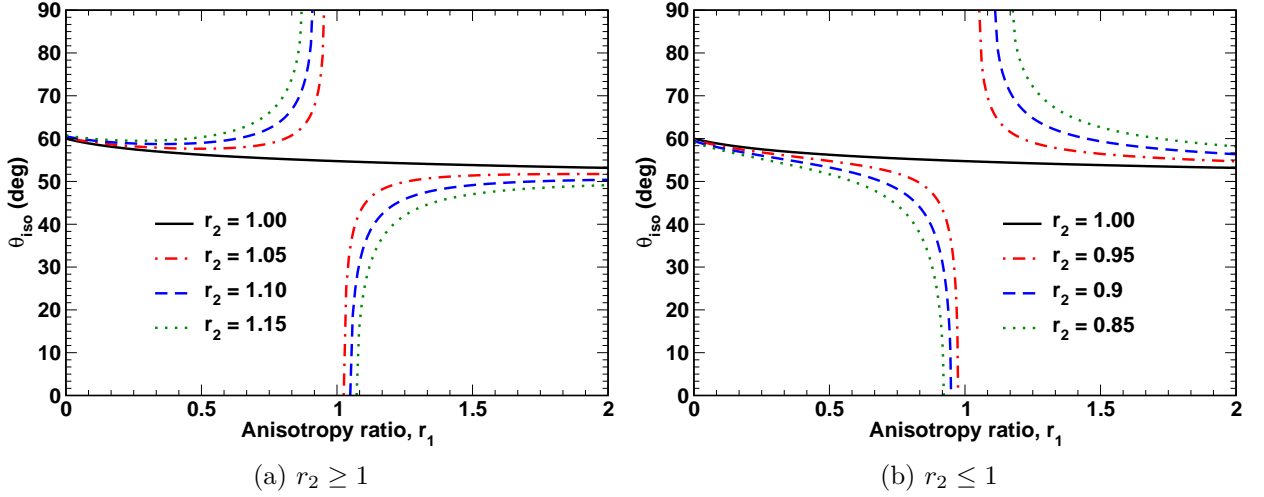

\centering
%\begin{center}
\subfloat[$r_2\ge 1$]{
%  \resizebox{80mm}{!}{\includegraphics*{theta_r1_gt1.eps}}
  \resizebox{80mm}{!}{\includegraphics*{fig8a.eps}}
\label{subfig:theta_r1_g1}
}
\subfloat[$r_2\le 1$]{
%  \resizebox{80mm}{!}{\includegraphics*{theta_r1_lt1.eps}}
  \resizebox{80mm}{!}{\includegraphics*{fig8b.eps}}
\label{subfig:theta_r1_l1}
}
%\end{center}
\caption{%
(Color online)
Isotropization tilt angle
versus the anisotropy ratio $r_1=\epsilon_1/\epsilon_{\perp}$
at different values of the biaxiality ratio $r_2=\epsilon_2/\epsilon_{\perp}$.
}
\label{fig:theta_r1}
\end{figure*}

We can now use Eq.~\eqref{eq:epsilon_ph}
and write down the condition of zero-field
isotropy~\eqref{eq:isotropy-1}
in the following explicit form:
   \begin{align}
&
\label{eq:isotropy-2}
r_2 \sqrt{u}-
\sqrt{r_2} =
(u_1-u+1)
\left(
\frac{r_2-1}{\sqrt{u}+\sqrt{r_2}}+u^{-1/2}
\right).
    \end{align}
The case of a uniaxially anisotropic FLC with $r_2=1$
can be treated analytically.
In this case, it is not difficult to check that $r_1=1$ gives 
the special solution of  Eq.~\eqref{eq:isotropy-2}
that does not depend on the tilt angle
and corresponds to an isotropic material with $r_1=r_2=1$.
Another solution is given by the relation
    \begin{align}
      \label{eq:theta-iso-uni}
      \sin^2 \theta_{\ind{iso}}=\frac{1}{2}+\frac{\sqrt{9+8 u_1}-3}{8 u_1}
    \end{align}
linking the isotropization angle $\theta_{\ind{iso}}$
and the anisotropy parameter, $u_1=r_1-1$.
In Fig.~\ref{fig:theta_r1}, this solution is represented by 
the solid line curve. The isotropization angle is shown
to be a slowly decreasing function of the anisotropy ratio $r_1$.
From Eq.~\eqref{eq:theta-iso-uni}, it starts from the maximal value of
$\theta_{\ind{iso}}$ is $\pi/3$ and decays approaching 
$\pi/4$.

When the biaxiality ratio $r_2$ differs from unity,
the solution of the isotropy condition~\eqref{eq:isotropy-2}
can only be written in the parametrized form as follows 
    \begin{align}
&
      \label{eq:theta-iso-r1}
      \begin{cases}
        \sin^2\theta_{\ind{iso}}=(u-1)(R_1(u)-1)^{-1},\\
        r_1=R_1(u),
      \end{cases}
\end{align}
where
    \begin{align}
&
    \label{eq:R1}
R_1(u)/\sqrt{u}=\sqrt{u} +
(\sqrt{u}+\sqrt{r_2})\frac{\sqrt{r_2 u}-1}{\sqrt{r_2 u}+1}.
    \end{align}
The $\theta_{\ind{iso}}$ versus $r_1$ curves 
computed from the representation~\eqref{eq:theta-iso-r1}
are shown in Fig.~\ref{fig:theta_r1}.
It can be seen that,
by contrast to the case of uniaxial anisotropy [$r_2=1$],
for biaxial FLCs with $r_2\ne 1$,
each curve has two branches separated by a gap.
The isotropization angle vanishes, $\theta_{\ind{iso}}=0$,
at one of the endpoints of the gap, $r_1=\sqrt{r_2}$,
whereas the angle $\theta_{\ind{iso}}$ equals $\pi/2$
at the other endpoint. 
Thus, 
dependence of the isotropization angle
on the anisotropy ratio $r_1$
being smooth and continuous
for uniaxially anisotropic FLCs
is found to be splitted into two branches
when the FLC anisotropy is biaxial.
From Fig.~\ref{fig:theta_r1},
one of the branches 
with $u_1 u_2 >0$ is associated with 
the endpoint at $r_1=\sqrt{r_2}$
and lies below of the solid line curve
representing conventional FLCs.
For this branch, the angle $\theta_{\ind{iso}}$
decreases with the biaxiality ratio
reaching zero at $r_2=r_1^2$.

\begin{figure*}[!tbh]
\centering 
%  \resizebox{95mm}{!}{\includegraphics*{dn_alp_uni.eps}}
  \resizebox{95mm}{!}{\includegraphics*{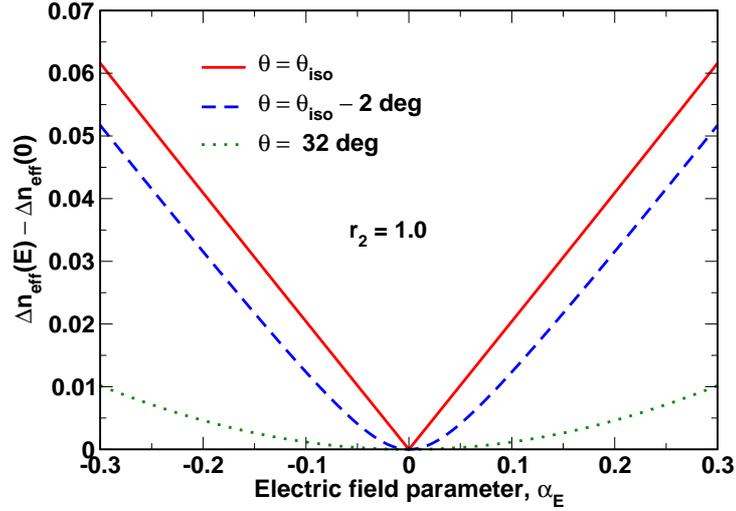}}
\caption{%
(Color online)
Electrically controlled birefringence,
$\Delta n_{\eff}(E)-\Delta n_{\eff}(0)$
[$\Delta n_{\eff}=n_{+}-n_{-}$],
versus the electric field parameter
at $r_1=1.32$ and $r_2=1$. 
}
\label{fig:dn_alp}
\end{figure*}

\begin{figure*}[!tbh]
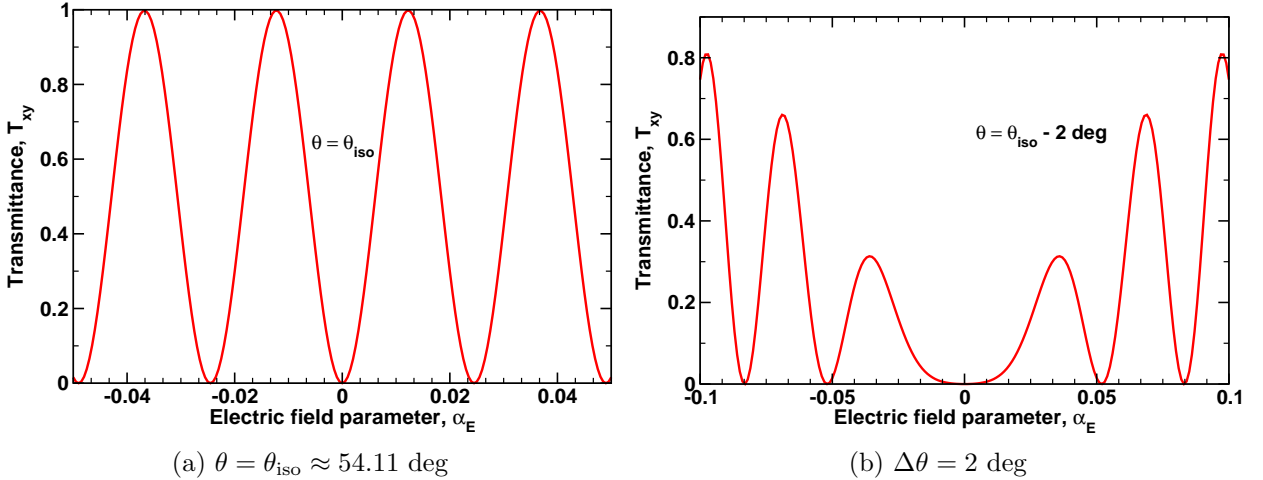

\centering
\subfloat[$\theta=\theta_{\ind{iso}}\approx 54.11$~deg]{
%  \resizebox{80mm}{!}{\includegraphics*{transmission_tc_uni.eps}}
  \resizebox{80mm}{!}{\includegraphics*{fig10a.eps}}
\label{subfig:transm_c_uni}
}
\subfloat[$\Delta\theta=2$~deg]{
%  \resizebox{80mm}{!}{\includegraphics*{transmission_tc-2_uni.eps}}
  \resizebox{80mm}{!}{\includegraphics*{fig10b.eps}}
\label{subfig:transm_c-2_uni}
}
\caption{%
(Color online)
Transmittance of light passing through crossed polarizers, 
$T_{xy}$, as a function of the electric
field parameter at (a)~$\theta=\theta_{\ind{iso}}\approx 54.11$~deg
and (b)~$\theta=52.11$~deg.
Parameters are listed in the caption of Fig.~\ref{fig:expt_th}. 
}
\label{fig:transmission_alpha}
\end{figure*}

We conclude this section with the remark
on the electro-optic response of DHFLC cells 
in the vicinity of the isotropization point.
One of the important factors governing the electric field dependence of
the transmittance~\eqref{eq:T-xy}
is the phase retardation $\delta$ (see Eq.~\eqref{eq:T-xy-approx})
proportional the effective birefringence $\Delta
n_{\eff}=n_{+}-n_{-}$.
The electrically dependent part of this birefringence 
is plotted as a function of the electric field parameter
in Fig.~\ref{fig:dn_alp}. 
It shown that, at $\theta=\theta_{\ind{iso}}$, the Kerr-like regime
breaks down and 
the birefringence is
dominated by the terms
linearly dependent on the electric field.
Such a Pockels-like behaviour manifests itself in 
the perfectly harmonic dependence of the transmittance,
$T_{xy}$, on the electric field parameter
depicted in Fig.~\ref{subfig:transm_c_uni}.

Figure~\ref{subfig:transm_c-2_uni} illustrates
the effect of small deviations from the isotropization angle.
Though the curve presented in Fig.~\ref{subfig:transm_c-2_uni}
and the ones for FLC-576A 
(see Fig.~\ref{fig:expt_th}) are quite similar in shape,
it is clear that sensitivity to the electric field 
and the magnitude of transmission peaks
are both considerably enhanced near the isotropization point.
Such behaviour comes as no surprise
and derives from the above discussed fact that 
this point plays the role of a singularity (an exceptional point)
(see Eqs.~\eqref{eq:cusp}--\eqref{eq:phase_jump}).

%%%%%%%%%%%%%%%%%%%%%%%
\section{Polarization-resolved angular patterns}
\label{sec:conoscopy}
%%%%%%%%%%%%%%%%%%%%%%

We can now combine
the general relations deduced in Sec.~\ref{sec:theory}
(and in Appendix~\ref{sec:planar}) using 
the transfer matrix method 
with 
the results of Sec.~\ref{sec:optics-dhf} 
to study 
the \textit{polarization-resolved angular (conoscopic)
patterns} 
describing the polarization structure behind the
conoscopic images
of short-pitch DHFLC cells that are characterized by
the effective dielectric tensor~\eqref{eq:elements-eff-diel-tensor}.
This polarization structure is represented by 
a two-dimensional (2D) distribution of
polarization ellipses and results from the interference of eigenmodes
excited in the  DHFLC cells by the plane waves with varying
direction of incidence.
Geometrically, the important elements
of  the 2D Stokes parameter fields 
are the polarization singularities
such as $C$ points (the points where the light wave is circular
polarized) and $L$ lines (the curves along which the polarization is
linear). 
In this section the focus of our attention will be on
the singularity structure of the polarization-resolved angular
patterns emerging after the DHFLC cells.
Our starting point is the computational method
used to evaluate the patterns as 
the polarization ellipse fields.

%%%%%%%%%%%%%%%%%%%%%%%
\subsection{Computational procedure}
\label{subsec:comput-proc}
%%%%%%%%%%%%%%%%%%%%%%

We shall use 
the electric field vector amplitudes 
of incident, reflected and transmitted waves 
conveniently rewritten in the circular basis
\begin{align}
 \label{eq:E_circ}
\vc{E}_{\alpha}^{(c)}=
\begin{pmatrix}
E_{+}^{(\alpha)}\\
E_{-}^{(\alpha)}
\end{pmatrix}
=
\mvc{C}
\begin{pmatrix}
E_{p}^{(\alpha)}\\
E_{s}^{(\alpha)}
\end{pmatrix},
\quad
 \mvc{C}=
\frac{1}{\sqrt{2}}
\begin{pmatrix}
1 & -i\\
1 & i
\end{pmatrix},
\quad
\alpha\in\{\inc,\transm,\refl\}.
\end{align}
and the incidence angles,
$\theta_{\ind{inc}}$ and $\phi_{\ind{inc}}$, 
related to
the lateral component of the wave
vector~\eqref{eq:k_p}
as follows
\begin{align}
  \label{eq:angles_inc}
  q_p=n_{\med}\sin\theta_{\inc},
\quad \phi_{p}=\phi_{\inc}, 
\end{align}
where $\theta_{\inc}$ ($\phi_{\inc}$)
is the \textit{polar (azimuthal) angle of incidence}.
Dependence of the polarization properties of the waves
transmitted through the DHFLC cell on 
the incidence angles, $\theta_{\ind{inc}}$ and $\phi_{\ind{inc}}$, 
will be of our primary concern. 

The transmission matrix describing
conoscopic patterns on 
the transverse plane of
projection
is given by~\cite{Kiselev:jpcm:2007,Kiselev:pra:2008}
\begin{align}
&
  \label{eq:T_con}
  \mvc{T}_{\ind{con}}(\rho,\phi)=\exp(-i\phi\bs{\sigma}_3)\,\mvc{T}_c(\rho,\psi_d-\phi)\,\exp(i\phi\bs{\sigma}_3)
\\
&
\label{eq:T_circ}
\mvc{T}_c(\rho,\psi_d-\phi)=
\begin{pmatrix}
  t_{++} & t_{+-}\\
t_{-+} & t_{--}
\end{pmatrix}
=
\mvc{C}\,\mvc{T}(q_p,\psi_d-\phi)\,\mvc{C}^{\dagger},
\\
&
\label{eq:rho-phi}
\rho= r\tan\theta_{\inc},
\quad
\phi=\phi_{\inc},
\quad
q_p/n_{\med}=\frac{\rho}{\sqrt{r^2+\rho^2}},
\end{align}
where $\rho$ and $\phi$ are the polar coordinates
in the observation plane
($x=\rho\cos\phi$ and $y=\rho\sin\phi$ are the Cartesian coordinates)
and $r$ is the aperture dependent scale factor.

The transmission matrix of DHFLC cells, 
$\mvc{T}(q_p,\psi_d-\phi)$, can be computed
from general formulas given in Appendix~\ref{sec:planar}
(see Eq.~\eqref{eq:TR_pln}).
For this matrix,
the parameters $\{\epsilon_z,\epsilon_{\parallel},\epsilon_{\perp},\psi\}$ 
that enter the expression for
the dielectric tensor of planar structures~\eqref{eq:diel-tensor-pln}
should be replaced with the characteristics
$\{\epsilon_z,\epsilon_{+},\epsilon_{-},\psi_d-\phi\}$
of the effective dielectric tensor~\eqref{eq:eff-diel-planar}
given in Eqs.~\eqref{eq:eff-diel-diag-planar}--~\eqref{eq:d_plus}.
In what follows the incident light is assumed to be linearly polarized
\begin{align}
&
\label{eq:E_inc_lin}
\vc{E}_{\inc}^{(c)}=
E_{\inc} \exp(-i\phi_p^{(\inc)}\bs{\sigma}_3)
\begin{pmatrix}
  1\\
1
\end{pmatrix},
\end{align}
where $\phi_p^{(\inc)}$ is the polarization azimuth of the incident
wave, and the state of polarization of 
the transmitted wave
\begin{align}
&
\label{eq:E_tr_con}
\vc{E}_{\transm}^{(c)}=
\begin{pmatrix}
  E_{+}^{(\transm)}\\
E_{-}^{(\transm)}
\end{pmatrix}
= \mvc{T}_{\ind{con}}(\rho,\phi)
\vc{E}_{\inc}^{(c)},
\end{align}
is defined by the polarization ellipse characteristics.
The orientation of the polarization ellipse is specified by 
the azimuthal angle of polarization (\textit{polarization azimuth})
\begin{align}
&
\label{eq:phi-transm}
2\phi_{p}^{(\transm)}
=\arg S\equiv \chi_s,
\quad
S=S_1+i S_2=2 \cnj{[E_{+}^{(\transm)}]} E_{-}^{(\transm)},
\end{align}
where $S_i$ is the $i$th component of the Stokes vector,
and its eccentricity is described by the signed ellipticity parameter
\begin{align}
&
\label{eq:ell-transm}
\epsilon_{\ind{ell}}^{(\transm)}=
\frac{|E_{+}^{(\transm)}|-|E_{-}^{(\transm)}|}{%
|E_{+}^{(\transm)}|+|E_{-}^{(\transm)}|}=
\tan\{2^{-1}\arcsin(S_3/S_0)\},
\quad
S_{3,\,0}=|E_{+}^{(\transm)}|^2\mp |E_{-}^{(\transm)}|^2
\end{align}
that will be referred to as the \textit{ellipticity}.
The ellipse is considered to be right handed
(left handed)
if its helicity is positive (negative),
so that $\epsilon_{\ind{ell}}^{(\transm)}>0$
($\epsilon_{\ind{ell}}^{(\transm)}<0$).

\begin{figure*}[!tbh]
%\vskip5mm
\centering
\subfloat[Star: $I_{C}=-1/2$ and $N_{C}=3$]{%
\resizebox{55mm}{!}{\includegraphics*{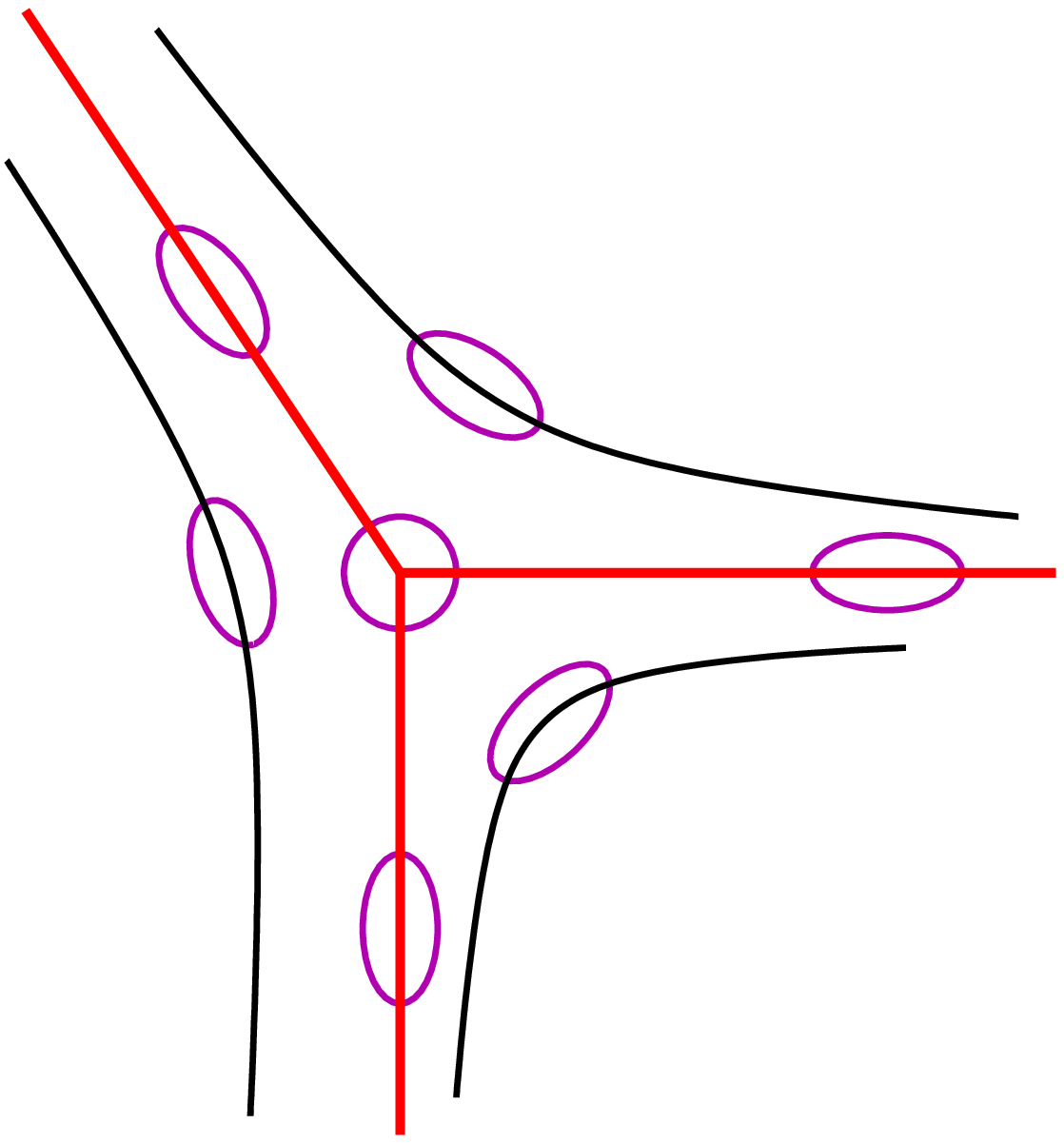}}
\label{fig:star}
}
\subfloat[Lemon: $I_{C}=+1/2$ and $N_{C}=1$]{%
\resizebox{55mm}{!}{\includegraphics*{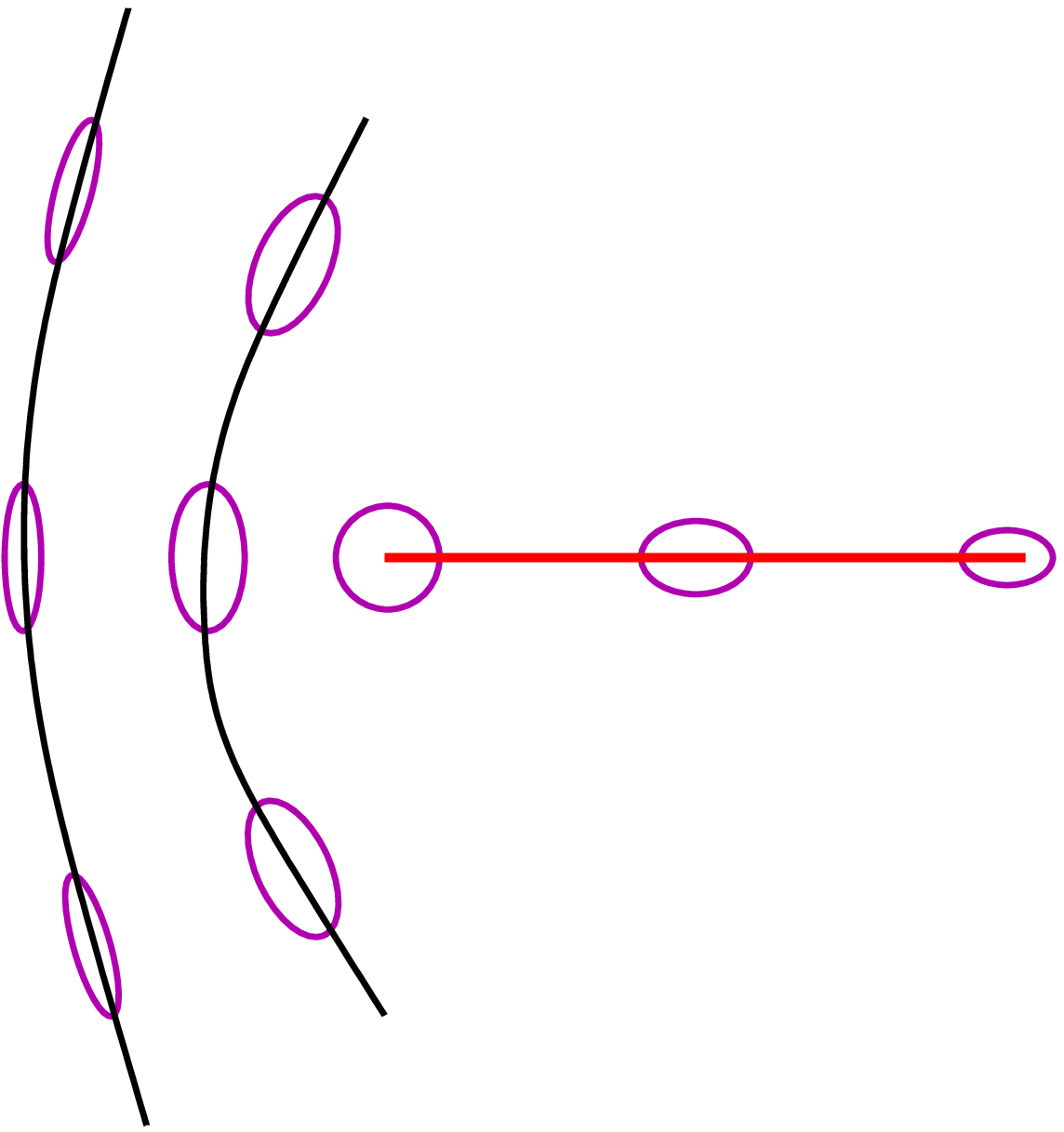}}
\label{fig:lemon}
}
\\
\subfloat[Monstar: $I_{C}=+1/2$ and $N_{C}=3$]{%
\resizebox{55mm}{!}{\includegraphics*{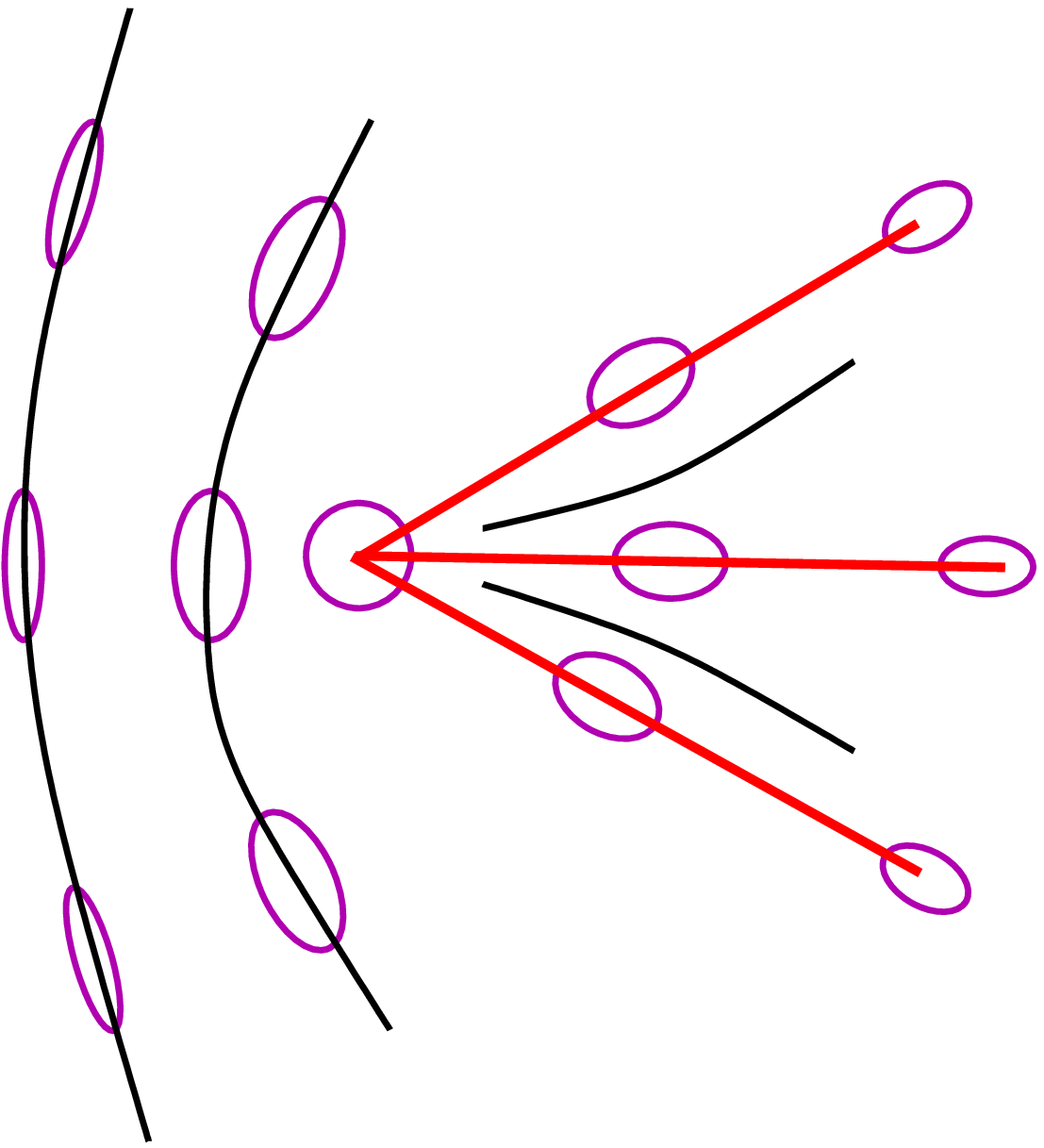}}
\label{fig:monstar}
}
\caption{%
(Color online)
Arrangement of the polarization ellipses around 
the $C$ points of three different types. } \label{fig:singul-types}
\end{figure*}

From Eq.~\eqref{eq:rho-phi},
the incidence angles and the points in the observation plane are in
one-to-one correspondence.
So, computing 
the polarization azimuth, $\phi_{p}^{(\transm)}$,
the ellipticity, $\epsilon_{\ind{ell}}^{(\transm)}$, 
at each point of the projection plane
yields the 2D field of polarization ellipses
which is called the polarization-resolved angular (conoscopic)
pattern.

The point where
$E_{\nu}^{(\transm)}=0$
and thus
the transmitted wave is circularly polarized
with $\epsilon_{\ind{ell}}^{(\transm)}=-\nu$
will be referred to as  the \textit{C$_{\nu}$ point}.

This is an example of
the \textit{polarization singularity}
that can be viewed as
the \textit{phase singularities} of the complex scalar field
$S=S_1+iS_2$
where the phase $\chi_{s}$
(see Eq.~\eqref{eq:phi-transm})
become indeterminate.
Such singularities are characterized by
the \textit{winding number} which is the signed number
of rotations  of the two-component field
$(S_1,S_2)$ around the circuit surrounding the
singularity~\cite{Merm:rpm:1979}.
The winding number also known as the \textit{signed strength of the
dislocation} is generically $\pm 1$.
 
Since the polarization azimuth~\eqref{eq:phi-transm} is defined modulo
$\pi$ and $2\phi_p=\arg S$, the dislocation strength is twice
the index of the corresponding $C_{\nu}$ point, $I_{C}$.
For generic $C$ points, $I_{C}=\pm 1/2$
and the topological index can be computed 
as the closed-loop contour integral
of the phase $\chi_s$ modulo $4\pi$
\begin{align}
  \label{eq:I_c-gen}
  I_{C}=\frac{1}{4 \pi}
\oint_{L}\dd\chi_s,
\end{align}
where $L$ is the closed path around the singularity.

In addition to the handedness and the index,
the $C$ points are classified according to the number of
streamlines, which are polarization lines 
whose tangent gives the polarization azimuth, 
terminating on the singularity.
This is the so-called \textit{line classification}
that was initially studied in the context of
umbilic points~\cite{Berry:jpa:1977}.
Mathematically,
the straight streamlines that terminate on the singularity
are of particular importance as they
play the role of separatrices, separating regions of streamlines
with differently signed curvature.
As is illustrated in Fig.~\ref{fig:singul-types},
for generic $C$ points,
the number of the straight lines, $N_{C}$, may either be 1 or 3.
This number is $3$ provided the index equals $-1/2$, $I_{C}=-1/2$,
and such $C$ points are called \textit{stars}.
At $I_{C}=1/2$, there are two characteristic patterns of polarization ellipses
around a $C$ point: (a)~\textit{lemon} with $N_{C}=1$ and
(b)~\textit{monstar} with $N_{C}=3$~\cite{Nye:prsl:1983a}.
Different quantitative criteria to distinguish between the $C$ points of
the lemon and the monstar types were deduced
in Refs.~\cite{Dennis:optcom:2002,Kis:jpcm:2007}.
From these criteria it can be inferred 
that a lemon becomes a monstar
as it approaches a star
and $C$ point annihilation occurs only between
stars and monstars~\cite{Dennis:optcom:2002,Dennis:optl:2008}.

The case of linearly polarized wave with 
$\epsilon_{\ind{ell}}^{(\transm)}=0$
provides another example of the polarization singularity
where the handedness is undefined.
The curves along which the polarization is linear are
called the \textit{L lines}.

From Eq.~\eqref{eq:phi-transm}
the $C$ points are  nodal points of 
the scalar complex function $S$
which can be found as intersection points of the 
Stokes parameter nodal lines $S_1=0$ and $S_2=0$.
Similarly, equation~\eqref{eq:ell-transm} implies that
nodes of the Stokes parameter field $S_3$
provide
the $L$ lines where $S_3=0$
and $\epsilon_{\ind{ell}}^{(\transm)}=0$.

\begin{figure*}[!tbh]
\centering
\subfloat[$\alpha_E=0$ and $\psi_d-\phi_p^{(\inc)} = 20$~deg]{
%  \resizebox{80mm}{!}{\includegraphics*{pol-f-00-00-20.eps}}
  \resizebox{80mm}{!}{\includegraphics*{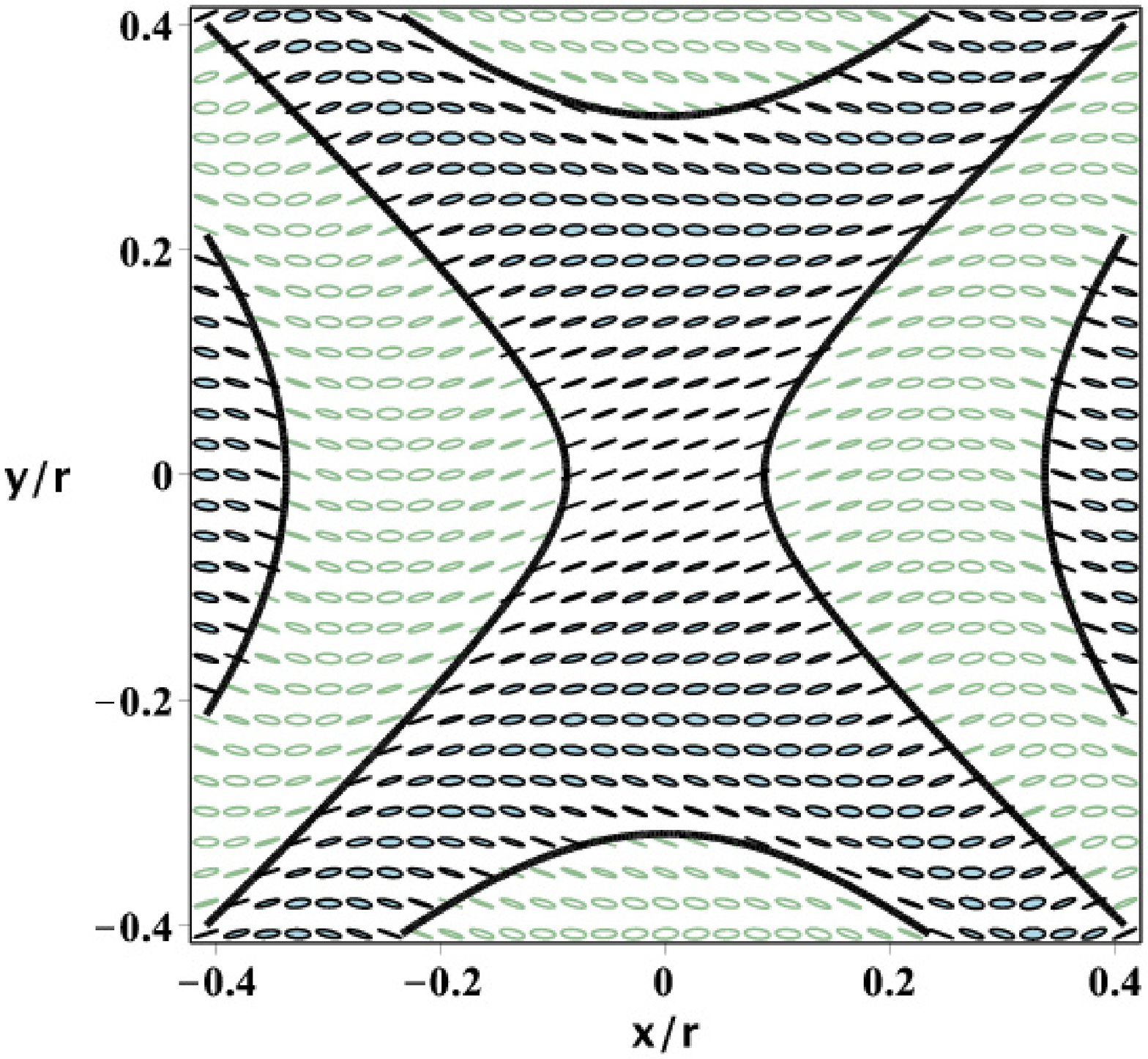}}
\label{subfig:pol-00-20}
}
\subfloat[$\alpha_E=0.3$ and $\psi_d-\phi_p^{(\inc)} = 20$~deg]{
  %\resizebox{80mm}{!}{\includegraphics*{pol-f-03-13-20.eps}}
  \resizebox{80mm}{!}{\includegraphics*{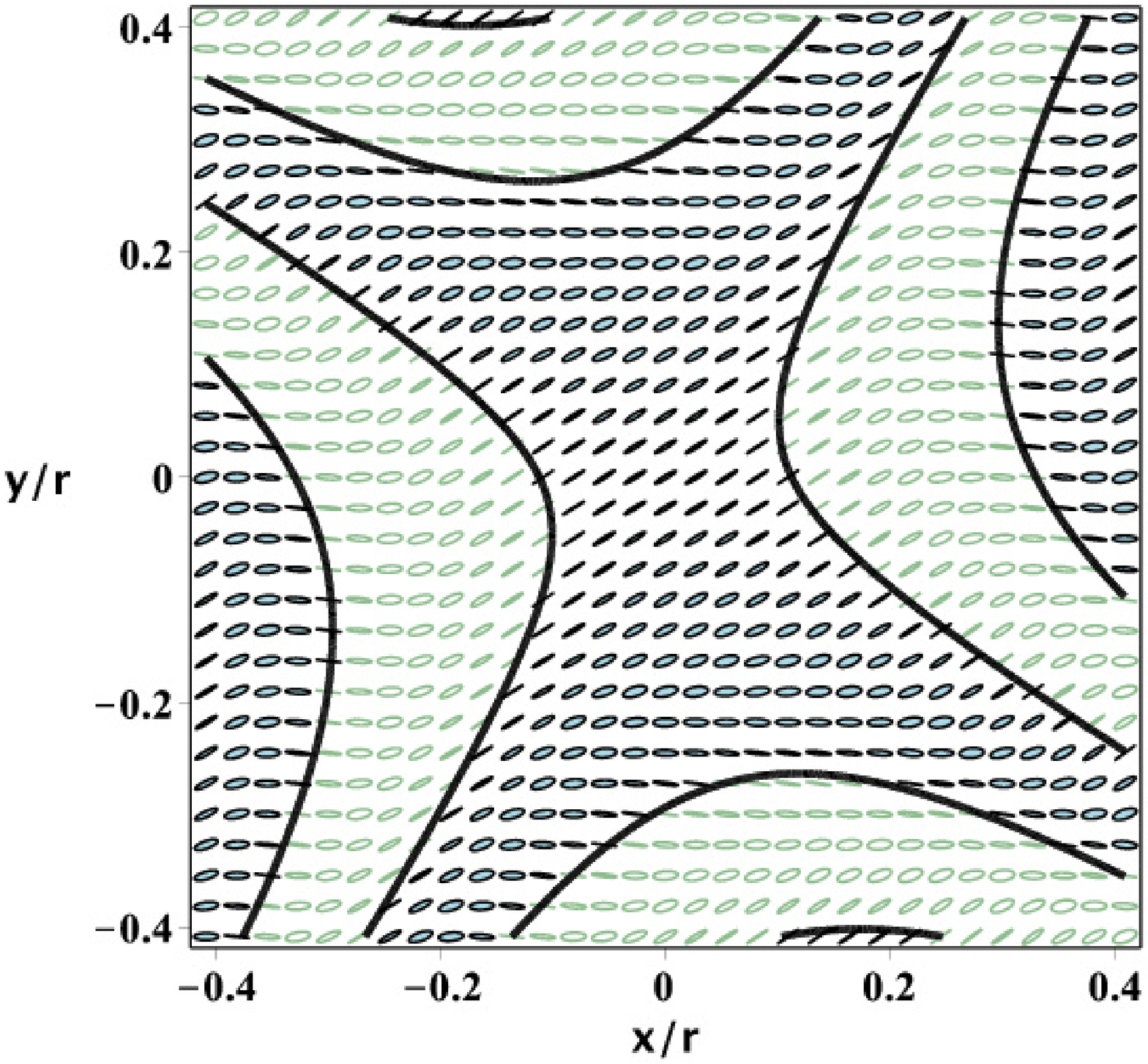}}
\label{subfig:pol-03-20}
}
\caption{%
(Color online)
Polarization-resolved conoscopic patterns computed
as polarization ellipse fields in the observation plane
for the DHFLC cell filled with the FLC mixture FLC-576A
(see Ref.~\cite{Kiselev:pre:2011} and the caption of
Fig.~\ref{fig:expt_th}).
Two cases are shown: (a)~$\alpha_E=0$
($n_{+}\approx 1.65$, $n_{-}=n_z\approx 1.532$
and $\psi_d=0$)
and (b)~$\alpha_E=0.3$
($n_{+}\approx 1.66$, $n_{-}\approx 1.529$
$n_z\approx 1.527$
and $\psi_d=13$~deg).
In both cases, the
angle between the in-plane optical axis
and the polarization plane of the incident light
is fixed at $\psi_d-\phi_p^{(\inc)} = 20$~deg.
$L$ lines are represented by thick black solid lines.
Left-handed and right-handed polarization is, respectively,
indicated by solid and open ellipses.
}
\label{fig:pol-f-20}
\end{figure*}

%%%%%%%%%%%%%%%%%%%%%%%
\subsection{Results}
\label{subsec:results}
%%%%%%%%%%%%%%%%%%%%%%

Now we
present  the theoretical results for the polarization-resolved
patterns of the DHFLC cells. 
These patterns are
computed for the cell 
of thickness $D=130$~\mum\ 
filled with the FLC mixture FLC-576A
which was studied in Ref.~\cite{Kiselev:pre:2011}
and described at the end of Sec.~\ref{subsec:kerr-effect}.

Our first remark is that
the angular dependence 
of the elements 
of the transmission matrix~\eqref{eq:T_circ}
is determined by
the angle difference
$\tilde{\phi}=\phi-\psi_d$
which is the angle
between 
the in-plane optical axis $\uvc{d}_{+}$
(see Eq.~\eqref{eq:d_plus})
and
the lateral wave vector $\vc{q}_p$
(see Eq.~\eqref{eq:k_p}).
Then the vector amplitudes
\begin{subequations}
 \label{eq:tilde_E_phi}
\begin{align}
&
  \label{eq:tilde_E_inc}
  \tilde{\vc{E}}_{\inc}^{(c)}=\exp(i\psi_d\bs{\sigma}_3)\vc{E}_{\inc}^{(c)}=
E_{\inc}\exp(-i\tilde{\phi}_p^{(\inc)}\bs{\sigma}_3)
\begin{pmatrix}
  1\\
1
\end{pmatrix},
\\
&
  \label{eq:tilde_E_trm}
\tilde{\vc{E}}_{\transm}^{(c)}=\exp(i\psi_d\bs{\sigma}_3)\vc{E}_{\transm}^{(c)}=
E_{\transm}\exp(-i\tilde{\phi}_p^{(\transm)}\bs{\sigma}_3)
\begin{pmatrix}
  1+\epsilon_{\ind{ell}}^{(\transm)}\\
1-\epsilon_{\ind{ell}}^{(\transm)}
\end{pmatrix},
\\
&
  \label{eq:tilde_phi_p}
\tilde{\phi}_p^{(\inc,\,\transm)}=\phi_p^{(\inc,\,\transm)}-\psi_d,
\end{align}
\end{subequations}
where
$\tilde{\phi}_p^{(\inc)}$
is the angle between 
the optical axis $\uvc{d}_{+}$
and
the polarization plane of the linearly polarized
incident wave
(see Eq.~\eqref{eq:E_inc_lin}),
describing the incident and  transmitted waves with
polarization ellipses rotated by the angle $\psi_p$
are related by the transformed transmission matrix
\begin{align}
&
  \label{eq:tilde-T_con}
  \tilde{\mvc{T}}_{\ind{con}}(\tilde{\phi})=
\exp(i\psi_d\bs{\sigma}_3)\,\mvc{T}_{\ind{con}}\,\exp(-i\psi_d\bs{\sigma}_3)=
\notag
\\
&
\exp[-i\tilde{\phi}\bs{\sigma}_3]\,\mvc{T}_{c}(\rho,-\tilde{\phi})\,
\exp[i\tilde{\phi}\bs{\sigma}_3].
\end{align}

From relation~\eqref{eq:tilde-T_con} it follows that, 
given the angle $\tilde{\phi}_p^{(\inc)}$,
the sole effect of
changing the azimuthal angle
of the optical axis: $\psi_d\to\psi_d+\Delta\psi$
is the rotation of the polarization ellipse field
by the angle $\Delta\psi$.
In DHFLC cells,
this effect 
manifests itself as 
the electric field induced rotation and
can be clearly seen in
Fig.~\ref{fig:pol-f-20} that
shows the patterns 
emerging after the DHFLC cell
calculated
at $\tilde{\phi}_p^{(\inc)}=-20$~deg
for two values of the electric filed parameter:
$\alpha_E=0$ 
(see Fig.~\ref{subfig:pol-00-20})
and $\alpha_E=0.3$ (see Fig.~\ref{subfig:pol-03-20}).

Figure~\ref{fig:pol-f-20} also illustrates 
the case of angular patterns 
that do not contain $C$ points.
 The geometry of such patterns is completely
characterized by the $L$ lines. 
Interestingly, 
at  $|\tilde{\phi}_p^{(\inc)}|>5$~deg,
it turned out that 
the $S_3$ nodal lines can be evaluated using
the simplified equation
\begin{align}
  \label{eq:L-line-approx}
  \sin\delta=0,
\quad
\delta=(q_e-q_o)h,
\end{align}
where $\delta$ is the phase retardation expressed in terms of
the eigenvalues of the matrix~\eqref{eq:Mij}
for uniformly anisotropic planar layers
(see Eqs.~\eqref{eq:qz_eo-pln-uni} and~\eqref{eq:qz_eo-pln-bia}),
that thus gives a sufficiently accurate approximation for 
$L$ lines.

The angle
$\tilde{\phi}_p^{(\inc)}$ 
can be regarded as the
governing parameter
whose magnitude 
determines the formation of $C$ points.
Given the aperture, the latter occurs only if the magnitude
of $\tilde{\phi}_p^{(\inc)}$ exceeds its critical value.

\begin{figure*}[!tbh]
\centering
\subfloat[$\alpha_E=0$ and $\psi_d-\phi_p^{(\inc)} = 39.5$~deg]{
 % \resizebox{80mm}{!}{\includegraphics*{pol-f-00-00-39_5.eps}}
  \resizebox{80mm}{!}{\includegraphics*{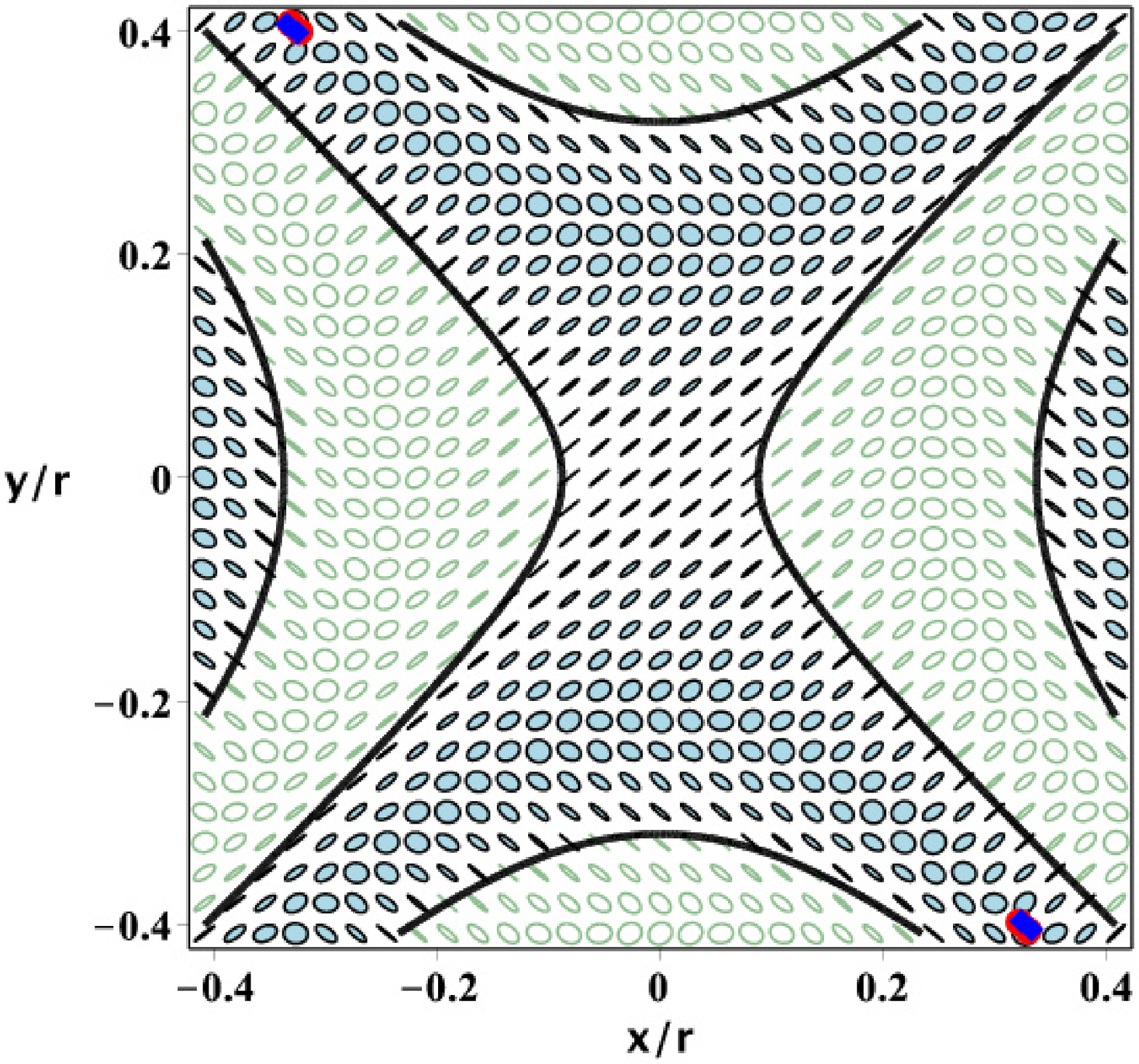}}
\label{subfig:pol-00-39_5}
}
\subfloat[$\alpha_E=0.3$ and $\psi_d-\phi_p^{(\inc)} = 39.5$~deg]{
%  \resizebox{80mm}{!}{\includegraphics*{pol-f-03-13-39_5.eps}}
  \resizebox{80mm}{!}{\includegraphics*{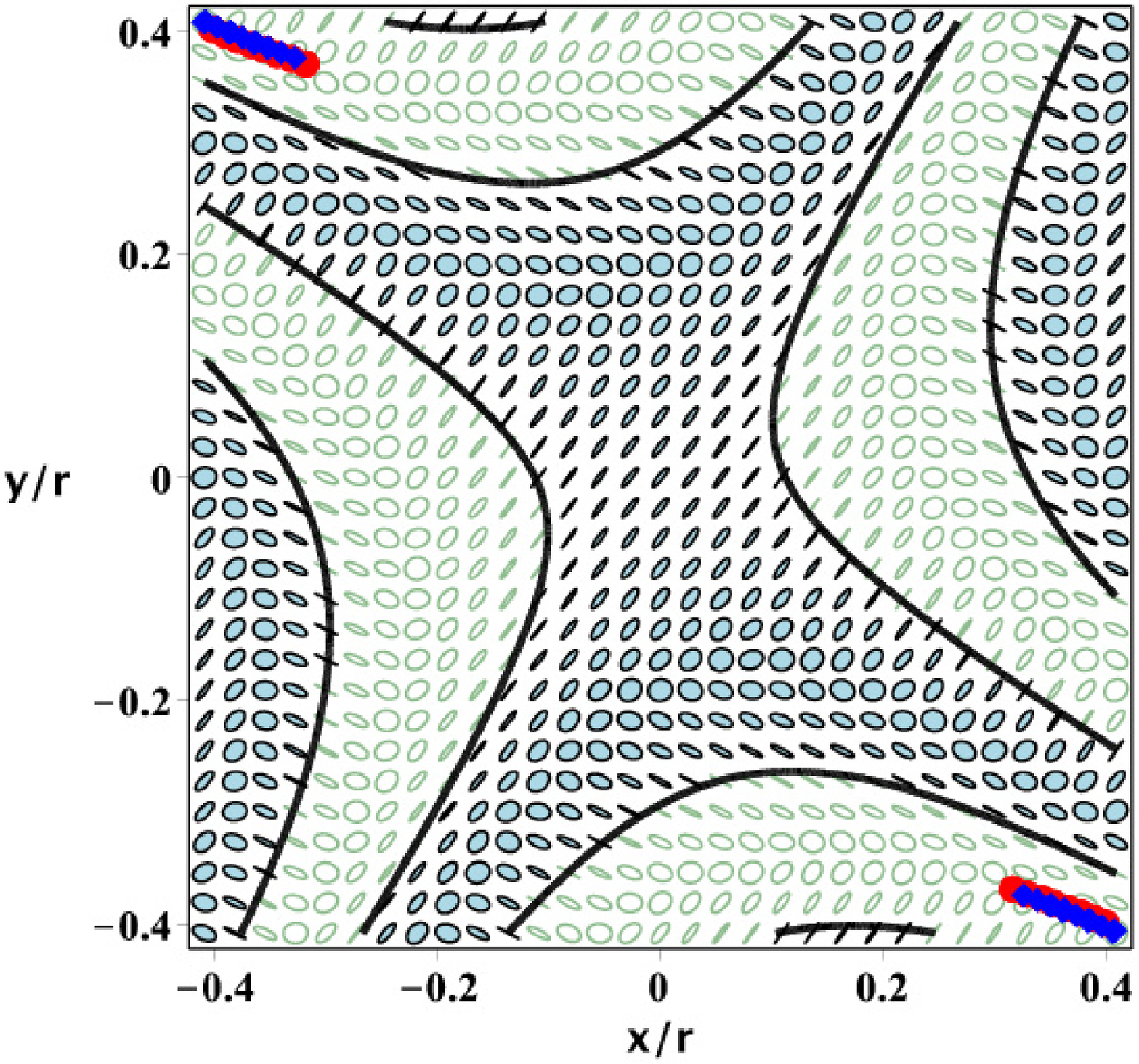}}
\label{subfig:pol-03-39_5}
}
\caption{%
(Color online)
Polarization-resolved conoscopic patterns computed
as polarization ellipse fields in the observation plane
for the DHFLC cell 
at $\psi_d-\phi_p^{(\inc)} = 39.5$~deg
(see the caption of
Fig.~\ref{fig:pol-f-20}).
Two cases are shown: (a)~$\alpha_E=0$
and (b)~$\alpha_E=0.3$.
$C$ points are indicated by
red circles (stars with $I_C=-1/2$)
and blue diamonds (monstars with $I_C=1/2$).
}
\label{fig:pol-f-39_5}
\end{figure*}

The case where the angle $\tilde{\phi}_p^{(\inc)}$
is close to the critical value
is illustrated in Fig.~\ref{subfig:pol-00-39_5}.
It can be seen that,
at $\tilde{\phi}_p^{(\inc)}=-39.5$~deg,
the singularity structure of
the polarization ellipse fields 
becomes complicated and is characterized by the presence of
symmetrically arranged star-monstar pairs
of $C$ points.
In addition to the above discussed electric-field-induced rotation,
the electric field is shown to facilitate the formation of $C$ points.
Clearly, the field induced biaxial anisotropy is responsible for 
this effect. 

\begin{figure*}[!tbh]
\centering
\subfloat[$\alpha_E=0$ and $\psi_d-\phi_p^{(\inc)} = 41$~deg]{
%  \resizebox{80mm}{!}{\includegraphics*{pol-f-00-00-41.eps}}
  \resizebox{80mm}{!}{\includegraphics*{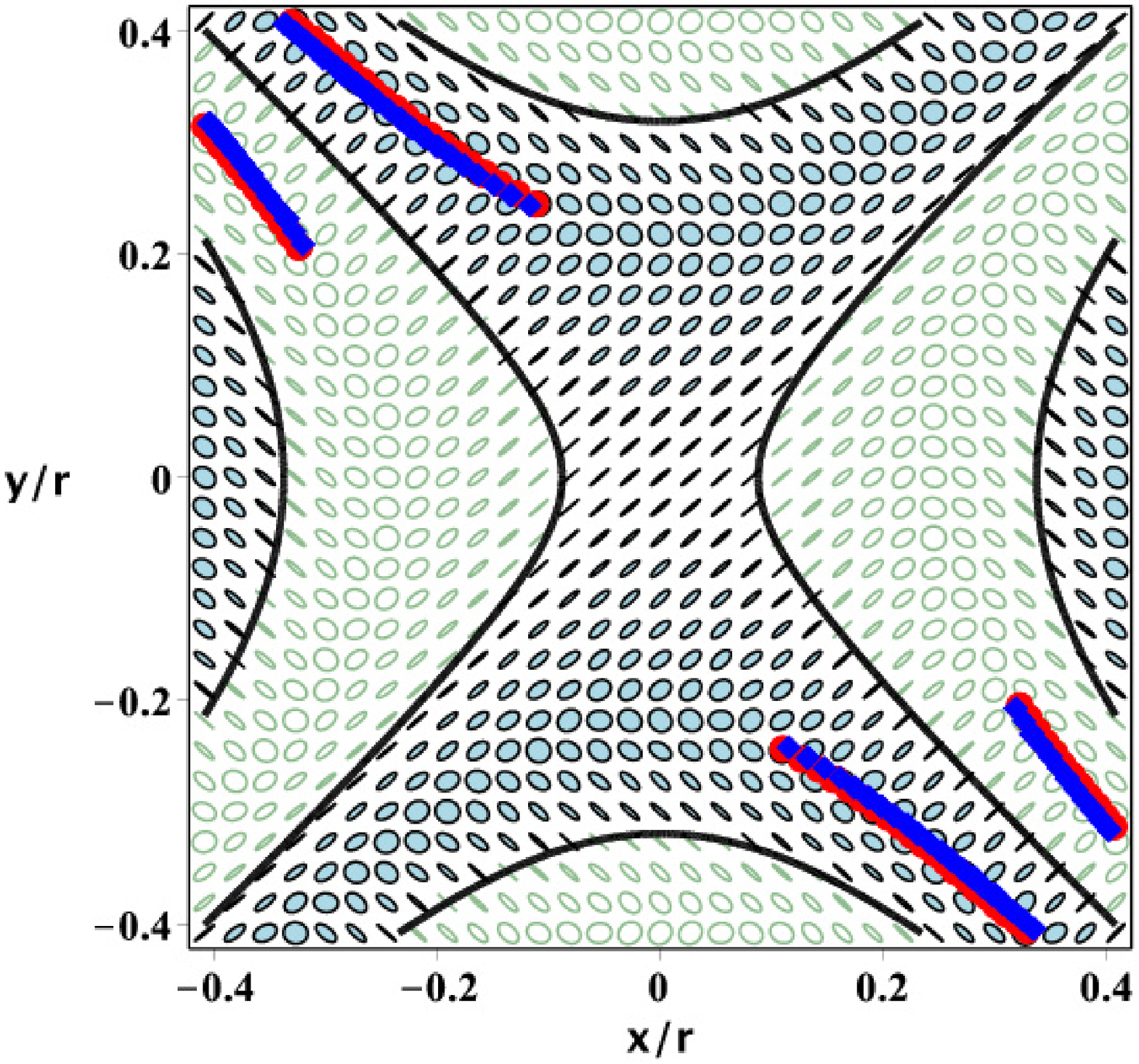}}
\label{subfig:pol-00-41}
}
\subfloat[$\alpha_E=0.3$ and $\psi_d-\phi_p^{(\inc)} = 41$~deg]{
%  \resizebox{80mm}{!}{\includegraphics*{pol-f-03-13-41.eps}}
  \resizebox{80mm}{!}{\includegraphics*{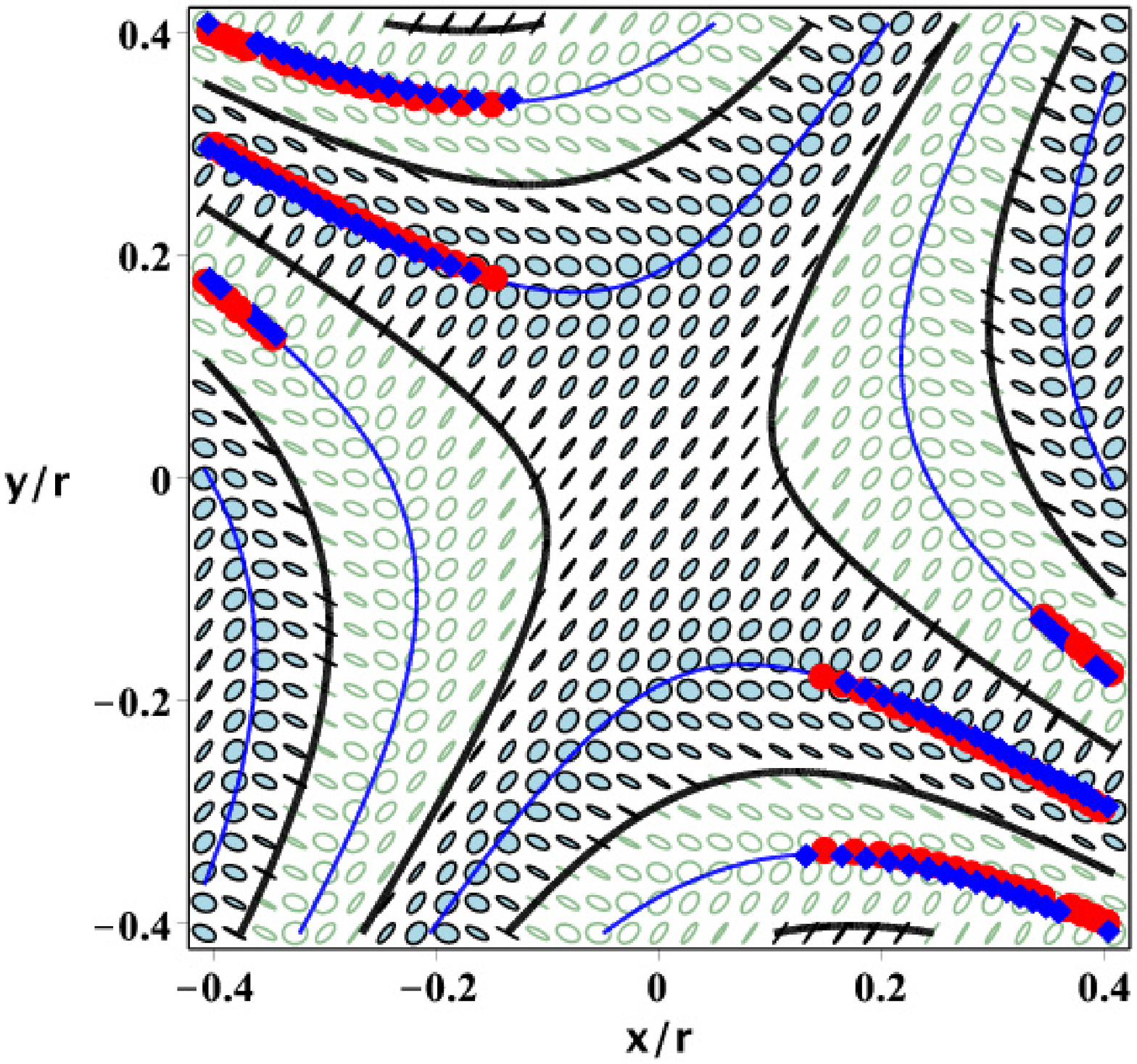}}
\label{subfig:pol-03-41}
}
\caption{%
(Color online)
Polarization-resolved conoscopic patterns computed
as polarization ellipse fields in the observation plane
for the DHFLC cell 
at $\psi_d-\phi_p^{(\inc)} = 41$~deg
(see the caption of
Fig.~\ref{fig:pol-f-20}).
Two cases are shown: (a)~$\alpha_E=0$
and (b)~$\alpha_E=0.3$.
$C$ points 
with $I_C=-1/2$ and $I_C=+1/2$
are indicated by
red circles (stars)
and blue diamonds (monstars), respectively.
Blue thin solid lines represent
the approximate $S_2$ nodal
lines computed by solving equation~\eqref{eq:S2-line-approx}.
}
\label{fig:pol-f-41}
\end{figure*}

% \begin{figure*}[!tbh]
% \centering
% \subfloat[$\alpha_E=0$ and $\psi_d-\phi_p^{(\inc)} = 40$~deg]{
%   \resizebox{80mm}{!}{\includegraphics*{pol-f-00-00-40.eps}}
% \label{subfig:pol-00-40}
% }
% \subfloat[$\alpha_E=0.3$ and $\psi_d-\phi_p^{(\inc)} = 40$~deg]{
%   \resizebox{80mm}{!}{\includegraphics*{pol-f-03-13-40.eps}}
% \label{subfig:pol-03-40}
% }
% \caption{%
% Polarization-resolved conoscopic patterns computed
% as polarization ellipse fields in the observation plane
% for the DHFLC cell 
% at $\psi_d-\phi_p^{(\inc)} = 40$~deg
% [see the caption of
% Fig.~\ref{fig:pol-f-20}].
% Two cases are shown: (a)~$\alpha_E=0$
% and (b)~$\alpha_E=0.3$.
% $C$ points 
% with $I_C=-1/2$ and $I_C=+1/2$
% are indicated by
% red circles (stars)
% and blue diamonds (monstars), respectively.
% }
% \label{fig:pol-f-40}
% \end{figure*}

In Fig.~\ref{fig:pol-f-41},
we show how the singularity structure 
of the polarization resolved patterns
develops
when the angle $\tilde{\phi}_p^{(\inc)}$ further increases.
This structure can be described as symmetrically arranged
chains of star-monstar pairs of $C$ points.
As is indicated in Fig.~\ref{subfig:pol-03-41},  
these tightly packed chains of $C$ points
are generally located in the vicinity
of the lines  
\begin{align}
  \label{eq:S2-line-approx}
  \cos\delta=0
\end{align}
that give a high accuracy approximation
for the $S_2$ nodal lines where $S_2=0$
and, similar to Eq.~\eqref{eq:L-line-approx},
are determined by the phase retardation
$\delta=(q_e-q_o)h$.
Note that, since 
$S_2-iS_3=2\cnj{[E_{s}^{(\transm)}]} E_{p}^{(\transm)}$,
applicability of  approximate formulas~\eqref{eq:L-line-approx}
and~\eqref{eq:S2-line-approx}
implies that the phase difference between
the components of the transmitted waves,
$E_{p}^{(\transm)}$ and $E_{s}^{(\transm)}$,
is close to the phase retardation:
$\arg\{\cnj{[E_{s}^{(\transm)}]} E_{p}^{(\transm)}\}\approx \delta$.
In Sec.~\ref{subsec:kerr-effect},
a similar approximation has been used to derive
the expression for the transmittance given by
Eq.~\eqref{eq:T-xy-approx}.

As it was mentioned in the previous section,
the loci of $C$ points on the projection plane
are determined by
intersections of the $S_2$ and $S_1$ nodal lines.
In our case, the star-monstar pairs are produced as a result
of small-scale oscillations of the nodal line
around the smooth curve described by Eq.~\eqref{eq:S2-line-approx}.
Experimentally, it is a challenging task
to resolve accurately 
the chains of $C$ points resulting from
such ripplelike oscillations
in polarimetry measurements.

The 2D polarization-resolved patterns are centrally symmetric
being invariant under inversion through the origin:
$(x,y)\to (-x,-y)$. The reason is that optical properties of planar anisotropic
structures are unchanged under
a 180-degree rotation about the normal to the cell
(the $z$ axis). More specifically,
we have the central symmetry relation
\begin{align}
  \label{eq:inv-pi}
    \tilde{\mvc{T}}_{\ind{con}}(\tilde{\phi})=  \tilde{\mvc{T}}_{\ind{con}}(\tilde{\phi}+\pi)
\end{align}
which is an immediate consequence of the fact that
the matrix $\mvc{M}$ given in
Eqs.~\eqref{eq:M12-pln} and~\eqref{eq:M21-pln}
remains intact when the azimuthal angle of the in-plane optic axis is changed  by 
$\pi$.

Another symmetry relation
\begin{align}
  \label{eq:refl}
    \tilde{\mvc{T}}_{\ind{con}}(-\tilde{\phi})=  \bs{\sigma}_1\tilde{\mvc{T}}_{\ind{con}}(\tilde{\phi})\bs{\sigma}_1
\end{align}
describes the transformation of the transmission
matrix~\eqref{eq:tilde-T_con}
under reflection in the mirror symmetry axis directed along $\uvc{d}_{+}$: 
$\tilde{\phi}\to -\tilde{\phi}$.
This relation immediately follows from Eq.~\eqref{eq:TR_refl_pln}
deduced in Appendix~\ref{sec:planar}.
By using formula~\eqref{eq:refl}, it is not difficult to show that
the polarization ellipse field 
$\{\tilde{\phi}_p^{(\transm)}(\rho,\tilde{\phi}),\epsilon_{\ind{ell}}^{(\transm)}(\rho,\tilde{\phi})\}$
transforms into its mirror
symmetric counterpart 
$
\{-\tilde{\phi}_p^{(\transm)}(\rho,-\tilde{\phi}),-\epsilon_{\ind{ell}}^{(\transm)}(\rho,-\tilde{\phi})\}
$
when
the polarization azimuth $\tilde{\phi}_p^{(\inc)}$ changes its sign:
$\tilde{\phi}_p^{(\inc)}\to - \tilde{\phi}_p^{(\inc)}$.

% It can be shown that the condition
% \begin{align}
% &
%   \label{eq:C_nu-loci}
%   t_{\nu\nu}(\rho,\tilde{\phi})
% =-t_{\nu,\,-\nu}(\rho,\tilde{\phi})\exp[2 i \nu
% (\tilde{\phi}-\tilde{\phi}_p^{(\inc)})]
% \end{align}
% gives 

%%%%%%%%%%%%%%%%%%%%%%%
\section{Discussion and conclusions}
\label{sec:discussion}
%%%%%%%%%%%%%%%%%%%%%%

 In this paper, we have performed
transfer matrix analysis of 
polarization-resolved angular patterns
emerging after 
electrically controlled
short-pitch DHFLC cells.
Our formulation of the transfer matrix method,
which is a suitably modified version of
the approach developed in 
Refs.~\cite{Kiselev:jpcm:2007,Kiselev:pra:2008,Kiselev:pre:2011},
involves the following steps:
(a)~derivation of the system of equations for
the tangential components of the wave field
in the $4\times 4$ matrix form (see Eq.~\eqref{eq:matrix-system});
(b)~introducing the evolution operator
(propagator)~\eqref{eq:evol-oprt} and
the scattering matrix~\eqref{eq:TR_gen};
(c)~defining the transfer matrix~\eqref{eq:W-op} 
through the propagator and, finally, 
(d)~deducing formulas~\eqref{eq:W_TR}
that link the transfer and scattering matrices.
Description of this method  
is augmented by discussion of a variety of unitarity and symmetry relations
(see Sec.~\ref{subsec:symmetry} and Appendix~\ref{sec:planar}),
with an emphasis on the special case
of anisotropic planar structures representing homogenized
DHFLC cells.
Interestingly, the relations
given in Eqs.~\eqref{eq:TR-sym-hom},~\eqref{eq:TR_hsym_pln}
and~\eqref{eq:TR_refl_pln} are shown to
be essentially independent of the assumption of lossless materials
and thus can be used when the medium is absorbing.
 
In general, 
we found that, owing to its mathematical structure, 
the transfer matrix approach 
provides the framework particularly useful
for in-depth analysis of symmetry related properties
(recent examples of such analysis can be found, e.g., in
Refs.~\cite{Altman:bk:2011,Dmitriev:ieee:2013,Luque:jqsrt:2014}).
Similarly, 
one of the important results of
a rigorous analysis performed within such a framework
in Ref.~\cite{Kiselev:pre:2011}
is the expression for
the effective dielectric tensor~\eqref{eq:eff-diel-tensor}
describing the electro-optical properties
of uniform lying FLC helical structures with
subwavelengh pitch.

In Sec.~\ref{sec:optics-dhf}
we have extended
theoretical considerations of Ref.~\cite{Kiselev:pre:2011}
to the case of biaxial FLCs
and have applied 
an alternative technique of averaging over distorted helix
to evaluate the dielectric tensor.
This technique is presented in Appendix~\ref{sec:FLC-spiral}
and gets around the difficulties of the method 
that relies on the well-known first-order
expression for a weakly distorted helix~\eqref{eq:Phi_weak}.
The modified averaging procedure 
allows high-order corrections to the dielectric tensor 
to be accurately estimated
and
improves agreement between the theory and the experimental
data in the high-field region (see Fig.~\ref{fig:expt_th}).

The resulting electric field dependence of  
the effective dielectric tensor~\eqref{eq:eff-diel-planar}
is linear (quadratic) for non-diagonal (diagonal) elements
with the coupling coefficients given by
Eq.~\eqref{eq:coupling-coeffs}.
These coupling coefficients along with the zero-field
dielectric constants~\eqref{eq:epsilon_ph}
determine how 
the applied electric field $E$ changes
the principal values of the effective dielectric tensor
(see Eqs.~\eqref{eq:epsilon_z} and~\eqref{eq:epsilon_pm})
and the azimuthal angle of optical axis~\eqref{eq:d_plus}.

Generally, at $E=0$, the DHFLC cell is uniaxially anisotropic
with the dielectric constants~\eqref{eq:epsilon_ph}, $\epsilon_p$ and
$\epsilon_h$, and 
the optical axis directed along the helix axis.
Then there are two most important effects induced by  the electric field: 
(a)~producing biaxial anisotropy by changing the eigenvalues
of the dielectric tensor; and 
(b)~rotation of in-plane optical axes by the field dependent angle $\psi_d$
defined in Eq~\eqref{eq:d_plus}.
At sufficiently low electric field $E$
and non-vanishing zero-field anisotropy,
the Kerr-like regime takes place so that
the principal values 
depend on the electric field quadratically
whereas the optical axis angle $\psi_d$
is approximately proportional to $E$.
This is the orientational Kerr effect previously
studied in Refs.~\cite{Kiselev:pre:2011,Kiselev:pre:2013,Kiselev:ol:2014}
for different geometries.

Our results on dependence of the coupling coefficients and the zero-field dielectric
constants on the smectic tilt angle $\theta$ described in
Sec.~\ref{subsec:tilt}
indicate a number of differences between uniaxial and biaxial FLCs.
What is more important, they show that the zero-field anisotropy
may vanish at certain value of $\theta$ which might be called
the isotropization angle:
$\theta=\theta_{\ind{iso}}$ (see Fig.~\ref{fig:eph_theta}).

In Sec.~\ref{subsec:isotropy},
the isotropization point
determined by the condition of zero-field
isotropy~\eqref{eq:isotropy-1}
is found to represent a singularity known as 
the exceptional point~\cite{Kato:bk:1995}.
For analytic continuation of
the dielectric tensor~\eqref{eq:eff-diel-planar}
in the complex $\alpha_E$ plane,
the exceptional points occur at the zeros
of the square root in Eq.~\eqref{eq:epsilon_pm}
where $[\Delta\epsilon]^2+[\gamma_{xy}\alpha_E]^2=0$.
In general, there are two pairs of complex conjugate values
of electric field parameter representing four exceptional (branch) points.
When the difference $\epsilon_h-\epsilon_p$ vanishes,
the two branch points coalesce on the real axis at the origin.

In the case of conventional uniaxial FLCs,
the analytic solution of the condition of zero-field isotropy 
can be obtained in the closed form and is given by
simple formula~\eqref{eq:theta-iso-uni}
where the isotropization angle, $\theta_{\ind{iso}}$, is found to 
be a decreasing function of the anisotropy parameter
$u_1=(\epsilon_1-\epsilon_{\perp})/\epsilon_{\perp}=r_1-1$.
For biaxial FLCs with $r_2=\epsilon_2/\epsilon_{\perp}\ne 1$, 
the solution can only be written
in the parametrized form~\eqref{eq:theta-iso-r1}.
As it can be seen in Fig.~\ref{fig:theta_r1},
the corresponding $\theta_{\ind{iso}}$-$r_1$ curves
are splitted into two branches separated by the gap.
These results significantly differ from the relation
$\cos^2\theta_{\ind{iso}}=(1+u_2/u_1)/3$
that can be easily obtained~\cite{Abdulhalim:apl:2012}
for the dielectric tensor~\eqref{eq:diel-tensor}
averaged over the FLC helix:
$\avr{\bs{\epsilon}}/\epsilon_{\perp}=[1
+(u_1\sin^2\theta+u_2)/2]\mvc{I}_3+
[u_1\cos^2\theta - (u_1\sin^2\theta+u_2)/2]\uvc{x}\otimes\uvc{x}$.
The difference stems from the fact
that, in our approach, the effective dielectric tensor
$\bs{\epsilon}_{\eff}$ is defined through 
the averaged differential propagation matrix
$\avr{\mvc{M}}$ and thus is not equal to the averaged dielectric
tensor~\eqref{eq:diel-tensor}:
$\bs{\epsilon}_{\eff}\ne \avr{\bs{\epsilon}}$

At the exceptional point,
the Kerr-like regime breaks down  
and  the electric field dependence of the birefringence 
becomes linear (see Fig.~\ref{fig:dn_alp}).
This might be called the Pockels-like regime
which is characterized by the harmonic electric field dependence
of the transmittance of light passing through crossed polarizers
(see Fig.~\ref{subfig:transm_c_uni}).
The curve shown in Fig.~\ref{subfig:transm_c-2_uni}
illustrates the electro-optical response of a DHFLC cell
near the exceptional point. It is seen  that 
sensitivity to the electric field and the magnitude of the
transmittance at peaks are both considerably enhanced
as compared to the case studied in Ref.~\cite{Kiselev:pre:2011}
(see also Fig.~\ref{fig:expt_th}).

We now try to put these results in a more general physical context.
In quantum physics, the exceptional points are known to produce
a variety of interesting phenomena
including level repulsion and crossing,
bifurcation, chaos and quantum 
phase 
transitions~\cite{Heiss:jpa:1990,Heiss:pre:2000,Song:prl:2010,Lee:pra:2012}.
For optical wave fields, a recent example is
unidirectional propagation (reflection) of 
light at the exceptional points in parity-time ($\mathcal{PT}$) symmetric periodic structures
and metamaterials that
has been the subject of intense studies~\cite{Lin:prl:2011,Yin:nmat:2013,Ming:pra:2014}.

To the best of our knowledge, the role of exceptional points in optics of liquid
crystal systems has yet to be recognized.
The main problem with conventional uniaxial FLCs is that 
the isotropy condition~\eqref{eq:isotropy-1} requires
large values of the smectic tilt angle
that are typically well above $50$~deg. 
Though there are no fundamental limitations
preventing preparation of FLC mixtures
with large tilt angles, this task still remains a challenge to deal with in
the future.  
Biaxial FLCs, where the isotropization tilt angle can be sufficiently small
when $\epsilon_1$ is close to $\sqrt{\epsilon_2}$,
also present a promising alternative approach for future work.

In Sec.~\ref{sec:conoscopy},
in order to gain further insight into
the electro-optical properties of the DHFLC cells, 
we have combined the transfer matrix approach
and the results for the effective dielectric tensor
to explore the polarization-resolved
angular patterns 
which are the polarization ellipse fields representing
the polarization structure of conoscopic images of DHFLC cells.
In the observation plane, such 2D patterns 
encode information on how the polarization state
of transmitted light is changed with the incidence angles
and exhibit singularities of a different kind,
the polarization singularities such as
$L$ lines (lines of linear polarization) and $C$ points (points of circular
polarization).
Note that, similar to the above discussed
exceptional point at which the angle $\psi_d$ becomes undetermined, 
$C$ points can be regarded as phase singularities 
(optical phase singularities are reviewed in Ref.~\cite{Dennis:progr_opt:2009}). 

Since the differential propagation matrix of planar structures
is invariant under rotation of in-plane optical axes by $\pi$,
the patterns are centrally symmetric [see Figs.~\ref{fig:pol-f-20}-~\ref{fig:pol-f-41}].
It was shown that,
at fixed the angle $\tilde{\phi}_p^{(\inc)}=\psi_d-\phi_p^{(\inc)}$
between the optical axis $\uvc{d}_{+}$
and the polarization plane of incident wave,
the sole effect of
the electric-field-induced rotation by the angle $\psi_d$
is rotation of the polarization ellipse field as a whole by the same angle. 

The symmetry axis of 
the $S_3$ nodal lines ($L$ lines) is found to be 
directed along $\uvc{d}_{+}$. 
When the $\tilde{\phi}_p^{(\inc)}$ is not too small,
they can be approximated by solving Eq.~\eqref{eq:L-line-approx}
and thus are mainly determined by the phase retardation $\delta$.
Similar remark applies to the $S_2$ lines 
and approximate formula~\eqref{eq:S2-line-approx}.

It turned out that this is the angle $\tilde{\phi}_p^{(\inc)}$
that plays the role of the parameter
governing formation of $C$ points.
When the magnitude of $\tilde{\phi}_p^{(\inc)}$
exceeds its critical value which, in our case, is close to $39$~deg,
$C$ points emerge as
symmetrically arranged and densely packed chains of star-monstar
pairs (see Figs.~\ref{fig:pol-f-39_5}-\ref{fig:pol-f-41}). 

So, in DHFLC cells, rotation of polarization ellipse fields and
formation of $C$ points are two most important effects
describing electrically induced transformations
of the polarization-resolved angular patterns.
These predictions can be verified experimentally.
This work is now in progress.

\begin{acknowledgments}
This work is supported by the HKUST grant ITP/039/12NP.
\end{acknowledgments}

\appendix

%%%%%%%%%%%%%%%%%%%%%%
\section{Operator of evolution}
\label{sec:op-evol}
%%%%%%%%%%%%%%%%%%%%%

We begin with the relation 
\begin{align}
  \label{eq:compos_law}
  \mvc{U}(\tau,\tau_0)=\mvc{U}(\tau,\tau_1)\cdot \mvc{U}(\tau_1,\tau_0)
\end{align}
known as the \textit{composition law}.
This result derives from the fact that the operator
$\mvc{U}(\tau,\tau_0)\cdot \mvc{U}^{-1}(\tau_1,\tau_0)$
is the solution of the system~\eqref{eq:evol_eq}
that satisfies the initial condition~\eqref{eq:evol_ic}
with $\tau_0$ replaced by $\tau_1$.

From the composition law~\eqref{eq:compos_law}
it immediately follow that 
the inverse of the evolution operator is given by
\begin{align}
  \label{eq:inv_evol_oper_1}
\mvc{U}^{-1}(\tau,\tau_0)=    
\mvc{U}(\tau_0,\tau)
\end{align}
and can be found by solving the initial value problem
\begin{align}
  \label{eq:eq_inv_evoper}
     i\pdrs{\tau}\mvc{U}^{-1}(\tau,\tau_0)=\mvc{U}^{-1}(\tau,\tau_0)\cdot
     \mvc{M}(\tau),
\quad
\mvc{U}^{-1}(\tau_0,\tau_0)=\mvc{I}_4.
\end{align}

For non-absorbing media with symmetric dielectric tensor, 
the matrix $\mvc{M}$ is real-valued, $\cnj{\mvc{M}}=\mvc{M}$,
and meets the following symmetry identities~\cite{Kiselev:pra:2008}:
\begin{align}
  \label{eq:symm_M}
  \cnj{\mvc{M}}=\mvc{M},\quad
\tcnj{(\mvc{G}\cdot\mvc{M})}=
\mvc{G}\cdot\mvc{M},
\quad
\mvc{G}=
\begin{pmatrix}
  \mvc{0}&\mvc{I}_2\\
\mvc{I}_2&\mvc{0}
\end{pmatrix},
\end{align}
where an asterisk and the superscript $T$ indicate 
complex conjugation and
matrix transposition, respectively. 
In this case, the evolution operator and its inverse
are related as follows: 
\begin{align}
  \label{eq:unit_U}
  \mvc{U}^{-1}(\tau,\tau_0)
=
\mvc{G}\cdot
\hcnj{\mvc{U}}(\tau,\tau_0)
\cdot
\mvc{G},
\end{align}
where a dagger will denote Hermitian conjugation.
By using the relations~\eqref{eq:symm_M},
it is not difficult to verify that
the operator on the right hand side of Eq.~\eqref{eq:unit_U}
is the solution of the Cauchy problem~\eqref{eq:eq_inv_evoper}.

%%%%%%%%%%%%%%%%%%%%%%%
\section{Uniformly anisotropic planar structures}
\label{sec:planar}
%%%%%%%%%%%%%%%%%%%%%%

In this section we present the results for
anisotropic planar structures characterized by
the dielectric tensor of the following form:
\begin{align}
  \label{eq:diel-tensor-pln}
  \epsilon_{ij}=\epsilon_z \delta_{ij} +
(\epsilon_{\parallel}-\epsilon_z)m_i m_j
+(\epsilon_{\perp}-\epsilon_z)l_i l_j,
\end{align}
where the optical axes
\begin{align}
  \label{eq:director_pln}
 \uvc{m}=(m_x,m_y,m_z)=(\cos\psi,\sin\psi,0),
\quad
\uvc{l}=\uvc{z}\times\uvc{m}=
 (-\sin\psi,\cos\psi,0)
\end{align}
lie in the plane of substrates (the $x$-$y$ plane).
The operator of evolution
can be expressed in terms of 
the eigenvalue and eigenvector matrices, 
$\mvc{\Lambda}\equiv\diag(\lambda_1,\lambda_3,\lambda_3,\lambda_4)$ and $\mvc{V}$, as follows
\begin{align}
  \label{eq:U-pln}
  \mvc{U}(h)=
     \exp\{i \mvc{M}\, h\}=\mvc{V}
      \exp\{i \mvc{\Lambda}\, h\}
\mvc{V}^{-1},
\quad
\mvc{M} \mvc{V}=\mvc{V} \mvc{\Lambda}.
\end{align}
For the dielectric tensor~\eqref{eq:diel-tensor-pln},
$\epsilon_{z \alpha}=\epsilon_{\alpha z}=0$
and $\mvc{M}_{ii}=0$ (see Eq.~\eqref{eq:Mij}).
Assuming that $\vc{q}_p=q_p\uvc{x}$,
we have
\begin{align}
&
  \label{eq:M12-pln}
  \mvc{M}_{12}=\mu
  \begin{pmatrix}
    1-q_p^2/n_z^2 & 0\\
0& 1
  \end{pmatrix},
\\
&
  \label{eq:M21-pln}
  \mu \mvc{M}_{21}=n_o^2
  \begin{pmatrix}
    1-u_a m_x^2 & -u_a m_x m_y\\
-u_a m_x m_y& 1-u_a m_y^2-q_p^2/n_o^2
  \end{pmatrix},
\end{align}
where $n_z=\sqrt{\mu \epsilon_z}$ and
$n_o=\sqrt{\mu \epsilon_{\perp}}$ are the principal refractive indices;
$u_a=(\epsilon_{\parallel}-\epsilon_{\perp})/\epsilon_{\perp}$
is the parameter of in-plane anisotropy.
For the case where the diagonal block-matrices, $\mvc{M}_{11}$ and
$\mvc{M}_{22}$, vanish,
it is not difficult to show that the eigenvector and eigenvalue
matrices can be taken in the following form:
\begin{align}
&
\label{eq:V-Q-pln}
\mvc{V}=
\begin{pmatrix}
            \mvc{E} & \mvc{E}\\
             \mvc{H} & -\mvc{H}
\end{pmatrix},
\quad
\mvc{\Lambda}=\diag(\mvc{Q},-\mvc{Q}),
\quad
\mvc{Q}=\diag(q_e,q_o).
\end{align}
In addition, the eigenvectors satisfy 
the orthogonality conditions 
(a proof can be found, e.g., in Appendix A of Ref.~\cite{Kiselev:pra:2008})
that, for the eigenvector matrix of the form~\eqref{eq:V-Q-pln},
can be written as follows
\begin{align}
  \label{eq:V-orth-pln}
  \tcnj{\mvc{V}} \mvc{G} \mvc{V}=\diag(\mvc{N},-\mvc{N}),
\quad
\mvc{N}=\diag(N_e,N_o)=2 \tcnj{\mvc{E}} \mvc{H}.
\end{align}

Upon substituting Eqs.~\eqref{eq:V-Q-pln}-~\eqref{eq:V-orth-pln}
into Eq.~\eqref{eq:W-op}, some rather 
straightforward algebraic manipulations give the transfer matrix
\begin{align}
&
\label{eq:W-pln}
  \mvc{W}=N_\med^{-1}\diag(\mvc{I}_2,\bs{\sigma}_3)
\tilde{\mvc{W}} \diag(\mvc{I}_2,\bs{\sigma}_3),
\\
&
 \label{eq:tW-pln}
\tilde{\mvc{W}}=
\begin{pmatrix}
  \mvc{A}_{+} & \mvc{A}_{-}\\
\mvc{A}_{-} & \mvc{A}_{+}
\end{pmatrix}
\mvc{W}_\dd
\begin{pmatrix}
 \mvc{A}_{+}^{T} &  -\mvc{A}_{-}^{T} \\
 -\mvc{A}_{-}^{T} &  \mvc{A}_{+}^{T}
\end{pmatrix}
\\
&
\label{eq:Wd-pln}
\mvc{W}_\dd=
\begin{pmatrix}
  \mvc{W}_{-} & \vc{0}\\
\vc{0} & \mvc{W}_{+}
\end{pmatrix},
\quad
  \mvc{W}_{\pm}=\exp[\pm i \mvc{Q} h] \mvc{N}^{-1},
\\
&
  \label{eq:A_pm-pln}
  \mvc{A}_{\pm}=\mvc{E}_{\med} \mvc{H}\pm\mvc{H}_{\med} \mvc{E},
\end{align}
where $N_{\med}=2 q_{\med}/\mu_{\med}$.
From Eq.~\eqref{eq:tW-pln}, 
the block $2\times 2$ matrices of $\tilde{\mvc{W}}$
are given by
\begin{subequations}
\label{eq:tW_ij-pln}
\begin{align}
&
\label{eq:tW_11-pln}
\tilde{\mvc{W}}_{11}=
N_{\med}\mvc{W}_{11}=
\mvc{A}_{+} \mvc{W}_{-} \mvc{A}_{+}^{T}-\mvc{A}_{-} \mvc{W}_{+} \mvc{A}_{-}^{T},
\\
&
\label{eq:tW_22-pln}
\tilde{\mvc{W}}_{22}=
N_{\med}\bs{\sigma}_3 \mvc{W}_{22} \bs{\sigma}_3=
\mvc{A}_{+} \mvc{W}_{+} \mvc{A}_{+}^{T}-\mvc{A}_{-} \mvc{W}_{-} \mvc{A}_{-}^{T},
\\
&
\label{eq:tW_21-pln}
\tilde{\mvc{W}}_{21}=
N_{\med}\bs{\sigma}_3 \mvc{W}_{21}=
-\tcnj{\tilde{\mvc{W}}}_{12}=
-N_{\med}\tcnj{[\mvc{W}_{12} \bs{\sigma}_3]}=
\notag
\\
&
=\mvc{A}_{-} \mvc{W}_{-} \mvc{A}_{+}^{T}-\mvc{A}_{+} \mvc{W}_{+} \tcnj{\mvc{A}}_{-}.  
\end{align}
\end{subequations}

Finally, we can combine Eq.~\eqref{eq:W_ii_unit}
and Eq.~\eqref{eq:W_21_unit}
with Eq.~\eqref{eq:W-pln} to derive
the expressions for the transmission and reflection matrices
\begin{align}
&
\label{eq:TR_pln}
\mvc{T}_{+}\equiv\mvc{T}(q_p,\psi)=
N_{\med}\tilde{\mvc{W}}_{11}^{-1},
\quad
\mvc{R}_{+}\equiv\mvc{R}= 
\bs{\sigma}_3 \tilde{\mvc{W}}_{21} \tilde{\mvc{W}}_{11}^{-1}
\end{align}
describing the case where the incident wave is impinging
onto the entrance face of the layer, $z=0$.

As it can be seen from formulas~\eqref{eq:tW_ij-pln},
the symmetry relations~\eqref{eq:W_ij_hom} are satisfied
even if the dielectric constants
$\epsilon_{\perp}$, $\epsilon_{\parallel}$ and
$\epsilon_z$ are complex-valued.
So, the applicability range of identities~\eqref{eq:TR-sym-hom} 
includes lossy (absorbing) anisotropic materials
described by the dielectric tensor of the form given in Eq.~\eqref{eq:diel-tensor-pln}.

% Interestingly, when the eigenvalue matrix~\eqref{eq:Q_N-pln}
% is real so that $\hcnj{\mvc{W}}_{+}=\mvc{W}_{-}$
% and $\hcnj{\mvc{A}}_{\pm}=\tcnj{\mvc{A}}_{\pm}$, 
% close inspection of the expressions~\eqref{eq:tW_ij-pln}
% shows that, in addition to the symmetry relations
% for uniform anisotropy~\eqref{eq:W_ij_hom},
% the case of uniform planar structure
% is characterized by the following algebraic identities:
Interestingly, inverse of the transfer matrix,
$\mvc{W}^{-1}$,
can be obtained from formula~\eqref{eq:W-pln}
by changing sign of the thickness parameter $h$:
$h\to -h$. In formulas~\eqref{eq:tW_ij-pln},
this transformation interchanges the matrices
$\mvc{W}_{+}$ and $\mvc{W}_{-}$, so that
$\tilde{\mvc{W}}_{11}\leftrightarrow \tilde{\mvc{W}}_{22}$
and $\tilde{\mvc{W}}_{12}\leftrightarrow \tilde{\mvc{W}}_{21}$.
So, from Eq.~\eqref{eq:W-pln}, the block matrices of $\mvc{W}^{-1}$
are given by
\begin{subequations}
\label{eq:W_inv_pln}
\begin{align}
&
  \label{eq:Wii_inv_pln}
  \mvc{W}_{11}^{(-1)}=
\bs{\sigma}_3 \mvc{W}_{22} \bs{\sigma}_3,
\quad
  \mvc{W}_{22}^{(-1)}=
\bs{\sigma}_3 \mvc{W}_{11} \bs{\sigma}_3,
\\
&
  \label{eq:Wij_inv_pln}
  \mvc{W}_{12}^{(-1)}=
\bs{\sigma}_3 \mvc{W}_{21} \bs{\sigma}_3,
\quad
  \mvc{W}_{21}^{(-1)}=
\bs{\sigma}_3 \mvc{W}_{12} \bs{\sigma}_3.
\end{align}
\end{subequations}
From the other hand,
Eqs.~\eqref{eq:W_TR} and~\eqref{eq:inv_W_TR}
give
the transfer matrix and its inverse, respectively,
expressed in terms of the transmission and reflection matrices,
$\mvc{T}_{\pm}$ and $\mvc{R}_{\pm}$.
 These expressions can now be substituted into
Eq.~\eqref{eq:W_inv_pln} to yield the relations 
\begin{align}
  \label{eq:TR_hsym_pln}
  \mvc{T}_{+}=
\bs{\sigma}_3 \mvc{T}_{-} \bs{\sigma}_3,
\quad
  \mvc{R}_{+}=
\bs{\sigma}_3 \mvc{R}_{-} \bs{\sigma}_3
\end{align}
linking the transmission (reflection) matrix,
$\mvc{T}_{+}\equiv \mvc{T}$ ($\mvc{R}_{+}\equiv \mvc{R}$),
and its mirror symmetric counterpart
 $\mvc{T}_{-}$ ($\mvc{R}_{-}$).

In conclusion of this section, we consider how
the transmission and reflection matrices transform
under the reflection in the $x-z$ plane when the azimuthal angle $\psi$
changes its sign: $\psi \to -\psi$.
From Eqs.~\eqref{eq:M12-pln} and~\eqref{eq:M21-pln}, 
we have
\begin{align}
  \label{eq:M_ij-refl_pln}
  \mvc{M}_{ij}(-\psi)=\bs{\sigma}_3 \mvc{M}_{ij}(\psi)
  \bs{\sigma}_3.
\end{align}
By using Eq.~\eqref{eq:M_ij-refl_pln}
it is not difficult to deduce a similar relation for
the transfer matrix
\begin{align}
  \label{eq:W_ij-refl_pln}
  \mvc{W}_{ij}(-\psi)=\bs{\sigma}_3 \mvc{W}_{ij}(\psi)
  \bs{\sigma}_3
\end{align}
that can be combined with Eq.~\eqref{eq:TR_hsym_pln}
to yield the result for the transmission and reflection matrices
in the final form:
\begin{align}
  \label{eq:TR_refl_pln}
  \mvc{T}_{\pm}(-\psi)=
\bs{\sigma}_3 \mvc{T}_{\pm}(\psi) \bs{\sigma}_3=
\mvc{T}_{\mp}(\psi),
\quad
  \mvc{R}_{\pm}(-\psi)=
\bs{\sigma}_3 \mvc{R}_{\pm}(\psi) \bs{\sigma}_3=
\mvc{R}_{\mp}(\psi).
\end{align}
An important point is that, similar
to identities~\eqref{eq:TR-sym-hom},
the assumption of lossless (non-absorbing) medium
is not required to derive 
the symmetry relations~\eqref{eq:TR_hsym_pln} and Eq.~\eqref{eq:TR_refl_pln}.

%%%%%%%%%%%%%%%%%%%%%%%%%%%
\subsection{Uniaxial anisotropy}
\label{subsec:uniaxial}
%%%%%%%%%%%%%%%%%%%%%%%%%%%

For the case of uniaxially anisotropic structure
with $\epsilon_z=\epsilon_{\perp}$,
it is not difficult to 
find the expressions for the eigenvalues
that enter the eigenvalue matrix~\eqref{eq:V-Q-pln}
\begin{align}
  \label{eq:qz_eo-pln-uni}
  q_{e}=\sqrt{n_e^2-q_p^2(1+u_a m_x^2)},
\quad
  q_{o}=\sqrt{n_o^2-q_p^2},
\end{align}
where 
$n_o=\sqrt{\mu\epsilon_{\perp}}$
($n_e=\sqrt{\mu\epsilon_{\parallel}}$)
is the refractive index for ordinary (extraordinary)
waves and
$u_a=(\epsilon_{\parallel}-\epsilon_{\perp})/\epsilon_{\perp}$
is the anisotropy parameter. 
Similarly, after computing the eigenvectors,
we obtain the eigenvector matrix in the following form:
\begin{align}
&
\label{eq:EH-pln-uni}
\mvc{E}=\mu\,
\begin{pmatrix}
            m_x[1-q_p^2/n_o^2] & m_yq_o \\
             m_y & -m_xq_o
\end{pmatrix},
\quad
\mvc{H}=
\begin{pmatrix}
              m_x q_e & m_y n_o^2 \\
             m_y q_e & -m_x [n_o^2-q_p^2]
\end{pmatrix},
\\
&
\label{eq:N_oe-pln-uni}
N_{e}=\dfrac{2q_e\mu}{n_o^2}(n_o^2-q_p^2m_x^2),
\quad
N_{o}=2q_o\mu (n_o^2-q_p^2m_x^2).
\end{align}
Equations~\eqref{eq:qz_eo-pln-uni}-~\eqref{eq:N_oe-pln-uni} 
can now be substituted into the general expression
for the transfer matrix defined by
formulas~\eqref{eq:W-pln}-~\eqref{eq:tW_ij-pln}
so as to obtain the transmission and reflection matrices~\eqref{eq:TR_hsym_pln}.

%%%%%%%%%%%%%%%%%%%%%%%%%%%
\subsection{Biaxial anisotropy}
\label{subsec:biaxial}
%%%%%%%%%%%%%%%%%%%%%%%%%%%

For the general case of biaxial anisotropy,
the expressions for the eigenvalues
are more complicated than those for uniaxially anisotropic
layer (see Eq.~\eqref{eq:qz_eo-pln-uni}).
These can be written in the following form:
\begin{align}
  \label{eq:qz_eo-pln-bia}
  2 (q_{e,\,o}^2+q_p^2)=\Tr{\tilde{\mvc{M}}}\pm
\sqrt{
[\Tr{\tilde{\mvc{M}}}]^2-4\det{\tilde{\mvc{M}}} 
},
\end{align}
where the matrix $\tilde{\mvc{M}}$
is given by
\begin{subequations}
 \label{eq:tilde-M-pln}
\begin{align}
&
  \label{eq:tilde-M-biax}
\tilde{\mvc{M}}
=
\mvc{Rt}(-\psi)\cdot
[\mvc{M}_{21}\cdot\mvc{M}_{12}-q_p^2\,\mvc{I}_2]
\cdot\mvc{Rt}(\psi)=
\begin{pmatrix}
  \tilde{m}_{11} & \tilde{m}_{12}\\
\tilde{m}_{21} & \tilde{m}_{22}
\end{pmatrix},
\\
&
\label{eq:tilde-M-11}
      \tilde{m}_{11}=n_{e}^2-(u_a+u_z[1+u_a])\,q_p^2 m_x^2,
\quad
      \tilde{m}_{12}=(u_a+u_z[1+u_a])\, q_p^2 m_x m_y,
\\
&
\label{eq:tilde-M-22}
     \tilde{m}_{22}=n_{o}^2-u_z q_p^2 m_y^2, \quad
      \tilde{m}_{21}=u_z q_p^2 m_x m_y,\quad
u_z=(\epsilon_{\perp}-\epsilon_z)/\epsilon_z.
\end{align}
\end{subequations}
It can be readily checked that the result for uniaxial 
anisotropy~\eqref{eq:qz_eo-pln-uni}
is recovered from Eq.~\eqref{eq:qz_eo-pln-bia} 
as the limiting case where 
the parameter of out-of-plane anisotropy
$u_z$ is negligible and $\epsilon_z=\epsilon_{\perp}$.

For the eigenvector and normalization matrices,
$\mvc{V}$ and $\mvc{N}$,
given in Eq.~\eqref{eq:V-Q-pln} and Eq.~\eqref{eq:V-orth-pln},
respectively,
the results are
\begin{align}
&
\label{eq:EH-pln-bia}
 \mvc{E}=\mvc{M}_{12}\cdot\mvc{H}\cdot\mvc{Q}^{-1},
\quad
\mvc{H}=\mvc{Rt}(\phi_d)\cdot
\begin{pmatrix}
  \tilde{m}_{22}-q_{e}^2-q_p^2& -\tilde{m}_{12}\\
-\tilde{m}_{21} & \tilde{m}_{11}-q_{o}^2-q_p^2
\end{pmatrix},
\\
&
\label{eq:N-pln-bia}
\mvc{N}=
\diag(N_{e},N_{o})= 
2
\tcnj{\mvc{H}}\cdot\mvc{M}_{12}\cdot\mvc{H}\cdot\mvc{Q}^{-1}.
\end{align}
These relations along with
formulas~\eqref{eq:W-pln}--\eqref{eq:tW_ij-pln}
give the transfer matrix for biaxially anisotropic films
with two in-plane optical axes.

Before closing this section we briefly comment on 
the important special case of normal incidence
that occurs at $q_p=0$.
In this case, the matrices $\mvc{A}_{\pm}$ defined in Eq.~\eqref{eq:A_pm-pln}
can be written in the factorized form
\begin{align}
  \label{eq:Apm_norm_pln}
  \mvc{A}_{\pm}(\psi)=\mvc{Rt}(\psi)\cdot
\mvc{A}_{\pm}(0)=
\begin{pmatrix}
  m_x & -m_y\\
m_y& m_x
\end{pmatrix}
\cdot
\begin{pmatrix}
  \dfrac{\mu_{\med}\,n_e\pm \mu\, n_{\med}}{\mu_{\med}}&0\\
0& -n_o\dfrac{\mu_{\med}\,n_o\pm \mu\, n_{\med}}{\mu_{\med}}
\end{pmatrix},
\end{align}
where $\mvc{Rt}(\phi)=\begin{pmatrix}
  \cos\phi &-\sin\phi\\
\sin\phi & \cos\phi
\end{pmatrix}$ 
is the matrix describing rotation
about the $z$ axis by the angle $\phi$.
Substituting Eq.~\eqref{eq:Apm_norm_pln}
into  Eq.~\eqref{eq:tW_ij-pln}
gives the block matrices  
\begin{align}
  \label{eq:tWij_norm_pln}
  \tilde{\mvc{W}}_{ij}(\psi)=\mvc{Rt}(\psi)\cdot
\tilde{\mvc{W}}_{ij}(0)\cdot\mvc{Rt}(-\psi)
\end{align}
expressed as a function of the director azimuthal
angle $\psi$.

The result for the transmission and reflection matrices
\begin{align}
  \label{eq:TR_norm_pln}
  \mvc{T}_{\pm}(\psi)=\mvc{Rt}(\pm\psi)\cdot
\mvc{T}(0)\cdot\mvc{Rt}(\mp\psi),
\quad
  \mvc{R}_{\pm}(\psi)=\mvc{Rt}(\mp\psi)\cdot
\mvc{R}(0)\cdot\mvc{Rt}(\mp\psi),
\end{align}
where the diagonal matrices $\mvc{T}(0)$ and $\mvc{R}(0)$
describe the  case in which the director~\eqref{eq:director_pln} 
lies in the incidence plane, 
immediately follows from the relations~\eqref{eq:TR_pln}
and~\eqref{eq:TR_hsym_pln}.

%%%%%%%%%%%%%%%%%%%%%%%
\section{Averaging over FLC helical structures}
\label{sec:FLC-spiral}
%%%%%%%%%%%%%%%%%%%%%%

In Sec.~\ref{subsec:diel-tensor-dhf},
the effective dielectric tensor~\eqref{eq:eff-diel-tensor} of
a deformed helix FLC cell
is expressed in terms of the averages
given in Eqs.~\eqref{eq:eta} and~\eqref{eq:beta}.
In this apppendix, we describe
how to perform averaging over
the helix pitch without recourse to 
explicit formulas for the azimuthal angle
$\Phi$ the FLC director~\eqref{eq:director}.
We 
assume that the azimuthal angle
is a function of $x$, so that
the free energy density
can be written in the following form:
\begin{align}
&
 \label{eq:free-energy-density}
f=\frac{K}{2} (\pdrs{x}\Phi - q_0)^2+V_{E}(\Phi)=
U_{K} (\pdrs{\phi_0}\Phi - 1)^2 - U_{E}\cos\Phi
\\
&
 \label{eq:electric-poten}
V_{E}(\Phi)=-\sca{\vc{E}}{\vc{P}}=-E P_{s}\cos\Phi,
\\
&
\label{eq:energy-scales}
U_E=E P_{s},
\quad
U_{K}=\frac{K q_0^2}{2},
\quad
\phi_0=q_0 x,
\end{align}
where $q_0=2\pi/P_0$ is the free twist wave number.
Then the free energy functional per unit volume can be
written as the free energy density averaged over
the helix pitch  
\begin{align}
&
  \label{eq:avr-free-energy}
  F[\Phi]/V=\avr{f}_{x}\equiv \frac{1}{P}\int_0^P f \dd x.
\end{align}
The first integral of the stationary point (Euler-Lagrange) equation
\begin{align}
  \label{eq:EL-eq}
  K \pdrs{x}^2\Phi-\pdrs{\Phi}V_{E}(\Phi)=0 
\end{align}
is given by
\begin{align}
  \label{eq:first-integral}
  U_{K} [\pdrs{\phi_0}\Phi]^2 + U_{E}\cos\Phi=E.
\end{align}
Assuming that $E\ge U_E$ and $\pdrs{x}\Phi$ is non-negative,
equation~\eqref{eq:first-integral} can be recast into the
differential form 
\begin{align}
  \label{eq:dphi}
\frac{\sqrt{m}\,\dd\Phi}{\sqrt{1-m_E\cos\Phi}}=q_0 \dd x,
\quad
m\equiv U_K/E,
\quad
m_E\equiv U_E/E.
\end{align}
Integrating Eq.~\eqref{eq:dphi} over the period
yields the relation for the helix wave number 
\begin{align}
  \label{eq:pitch}
  \sqrt{m}\avr{R}_{\Phi}\equiv
\frac{\sqrt{m}}{2\pi}\int_0^{2\pi}\frac{\dd
  \Phi}{\sqrt{1-m_E\cos\Phi}}=
q_0/q,
\quad
q= 2\pi/P,
\end{align}
where $\avr{\dots}_{\Phi}=(2\pi)^{-1}\int_0^{2\pi}\ldots\dd \Phi$.
This relation gives the helix pitch, $P$,
expressed in terms of the dimensionless parameter $\tau=U_K/E$.
We can now use equation~\eqref{eq:pitch} to rewrite
Eq.~\eqref{eq:dphi} in the following form 
\begin{align}
  \label{eq:dphi-2}
  \frac{R\,\dd\Phi}{\avr{R}_{\Phi}}=\dd \phi,
\quad
\phi= q x,
\quad
R=[1-m_E \cos\Phi]^{-1/2}.
\end{align}
An important consequence of
this equation is the relation
\begin{align}
  \label{eq:avr-phi-Phi}
  \avr{\dots}_{\phi}=\avr{R \dots}_{\Phi}/\avr{R}_{\Phi}
\end{align}
that allows to perform averaging over the helix pitch
by computing integrals over the azimuthal angle $\Phi$.
In particular, with the help of
Eqs.~\eqref{eq:avr-phi-Phi} and~\eqref{eq:pitch},
it is not difficult to deduce the following expression
for the free energy~\eqref{eq:avr-free-energy}:
\begin{align}
  \label{eq:avr-f}
  \avr{f}_{\phi}=U_K(1-2 q/q_0+m^{-1} [1 - 2m_E
  \avr{\cos\Phi}_{\phi}])=
\notag
\\
U_K\left[
1-m^{-1}\bigl\{
1 + 2 (\sqrt{m}-\avr{R^{-1}}_{\Phi})/\avr{R}_{\Phi}
\bigr\}
\right].
\end{align}

In the low voltage regime,
the parameter $m_E$ is small
and the left hand side of Eq.~\eqref{eq:dphi-2}
can be expanded into the power series in $m_E$.
The expansion up to the second order terms
is given by
\begin{align}
  \label{eq:Z-approx}
  R/\avr{R}_{\Phi}\approx 1 +\frac{m_E}{2}\cos\Phi
+\frac{3 m_E^2}{16}\cos (2\Phi)
\end{align}
and can be used to average
the $z$ component
of the polarization vector $\vc{P}_s$
defined in Eq.~\eqref{eq:pol-vector}.
The result reads
\begin{align}
  \label{eq:avr-cos}
  \avr{\cos\Phi}_{\phi}=
\avr{R \cos\Phi}_{\Phi}/\avr{R}_{\Phi}
\approx m_E/4=\chi_{E} E/P_s\equiv 
\alpha_E,
\end{align}
where $\chi_E=\partial \avr{P_z}/\partial E$ is
the dielectric susceptibility of the Goldstone mode~\cite{Carlsson:pra:1990,Urbanc:ferro:1991} 
and $P_z=P_s\cos\Phi$.
Similarly, the averages that enter the formulas for the elements
of the effective dielectric tensor~\eqref{eq:epsilon-ij-planar}
can be expressed in terms of the electric field parameter
$\alpha_E$ as follows:
\begin{subequations}
  \label{eq:averages-planar}
\begin{align}
&
\avr{v_{zz}^{-1}}_{\phi}=
\avr{(1+v \sin^2\Phi)^{-1}}_{\phi}
\approx [1+v]^{-1/2}(1+3 v\gamma_v^2\alpha_E^2),
\\
&
\avr{v_{zz}^{-1}\cos^2\Phi}_{\phi}
\approx
\gamma_v(1+3 [1+v]^{1/2}\gamma_v\alpha_E^2),
\\
&
\avr{v_{zz}^{-1}\cos\Phi}_{\phi}
\approx
2 \gamma_v\alpha_{E},
\quad
\gamma_v= [\sqrt{1+v}+1]^{-1},
\\
  &
\avr{v_{zz}^{-1}\sin\Phi}_{\phi}=
\avr{v_{zz}^{-1}\sin\Phi\cos\Phi}_{\phi}=
0.
\end{align}
\end{subequations}
Substituting the relations~\eqref{eq:averages-planar}
into Eqs.~\eqref{eq:epsilon-ij-planar}
give the effective dielectric tensor~\eqref{eq:eff-diel-planar}
which is expressed in terms
of the zero-field dielectric constants
\begin{align}
&
\label{eq:epsilon_h-v}
\epsilon_h
/\epsilon_{\perp}
=1+(r_1/r_2-1-v)[(1+v)^{-1/2}+u_2\gamma_v],
\\
&
\label{eq:epsilon_p-v}
\epsilon_p
=\epsilon_2\sqrt{1+v},
\quad
\gamma_v=[\sqrt{1+v}+1]^{-1},
\end{align}
and the coupling coefficients
\begin{align}
  \label{eq:coupling-coeffs-v}
  &
\gamma_{xx}
/\epsilon_{\perp}=
3 (r_1/r_2-1-v)\gamma_v^2[v(1+v)^{-1/2}+u_2(1+v)^{1/2}],
\\
&
\gamma_{yy}
=3 \epsilon_2 v \gamma_v^2\sqrt{1+v},
\quad
% \\
% &
\gamma_{xy}
=2 (\epsilon_{1}-\epsilon_{\perp})\gamma_v\cos\theta\sin\theta.
\end{align}

%\bibliographystyle{apsrev}
%\bibliographystyle{apsrev4-1}
%\bibliographystyle{lc}
%\bibliography{optics,polymer,scatter,lc,quant,hk,flc,qft,math,my_papers}


\end{document}